\documentclass[sigplan,screen]{acmart}

\usepackage{graphicx}
\usepackage{textcomp}
\usepackage{xcolor}

\usepackage{multirow}
\usepackage{listings}
\usepackage{makecell}
\usepackage{mathtools}
\usepackage{subcaption}
\usepackage{comment}

\usepackage{algorithm}
\usepackage{algpseudocode}
\usepackage{pifont}

\newcommand{\Fig}[1]{Fig.~\ref{#1}}
\newcommand{\Tbl}[1]{Tbl.~\ref{#1}}
\newcommand{\Sec}[1]{Sec.~\ref{#1}}
\newcommand{\Alg}[1]{Alg.~\ref{#1}}

\newcommand{\revise}[1]{{\color{black}{#1}}}

\newcommand{\mymethod}{M$^{\text{2}}$XFP}  

\newcommand\blankfootnote[1]{%
            \let\thefootnote\relax\footnotetext{#1}%
            \let\thefootnote\svthefootnote%
          }

\usepackage{balance}

\copyrightyear{2026}
\acmYear{2026}
\setcopyright{cc}
\setcctype{by}
\acmConference[ASPLOS '26]{Proceedings of the 31st ACM International Conference on Architectural Support for Programming Languages and Operating Systems, Volume 2}{March 22--26, 2026}{Pittsburgh, PA, USA}
\acmBooktitle{Proceedings of the 31st ACM International Conference on Architectural Support for Programming Languages and Operating Systems, Volume 2 (ASPLOS '26), March 22--26, 2026, Pittsburgh, PA, USA}
\acmPrice{}
\acmDOI{10.1145/3779212.3790185}
\acmISBN{979-8-4007-2359-9/2026/03}

\begin{document}

\title{\mymethod{}: A Metadata-Augmented Microscaling Data Format for Efficient Low-bit Quantization}

\author{Weiming Hu}
\email{weiminghu@sjtu.edu.cn}
\orcid{0009-0003-5115-0498}
\affiliation{%
  \institution{Shanghai Jiao Tong University}
  \city{Shanghai}
  \state{}
  \country{China}}
\affiliation{%
  \institution{Shanghai Qi Zhi Institute}
  \city{Shanghai}
  \country{China}}

\author{Zihan Zhang}
\email{tiancaizhangdaxian@sjtu.edu.cn}
\orcid{0009-0008-7683-2934}
\affiliation{%
  \institution{Shanghai Jiao Tong University}
  \city{Shanghai}
  \country{China}}

\author{Haoyan Zhang}
\email{h.y.zhang-zdy@sjtu.edu.cn}
\orcid{0009-0009-8634-5395}
\affiliation{%
  \institution{Shanghai Jiao Tong University}
  \city{Shanghai}
  \country{China}}
\affiliation{%
  \institution{Shanghai Qi Zhi Institute}
  \city{Shanghai}
  \country{China}}

\author{Chen Zhang}
\email{chenzhang.sjtu@sjtu.edu.cn}
\orcid{0000-0003-2762-2726}
\affiliation{%
  \institution{Shanghai Jiao Tong University}
  \city{Shanghai}
  \country{China}}
\authornote{Corresponding authors.}

\author{Cong Guo}
\email{guocong@sjtu.edu.cn}
\orcid{0000-0002-4479-5525}
\affiliation{%
  \institution{Shanghai Jiao Tong University}
  \city{Shanghai}
  \country{China}}

\author{Yu Feng}
\email{y-feng@sjtu.edu.cn}
\orcid{0000-0002-2192-5737}
\affiliation{%
  \institution{Shanghai Jiao Tong University}
  \city{Shanghai}
  \country{China}}

\author{Tianchi Hu}
\email{hutianchi1@huawei.com}
\orcid{0009-0004-2986-9858}
\affiliation{%
  \institution{Computing Product Line, Huawei}
  \city{Shanghai}
  \country{China}}

\author{Guanglin Li}
\email{liguanglin10@huawei.com}
\orcid{0009-0000-8996-3775}
\affiliation{%
  \institution{Computing Product Line, Huawei}
  \city{Shanghai}
  \country{China}}

\author{Guipeng Hu}
\email{huguipeng@huawei.com}
\orcid{0009-0007-7721-0048}
\affiliation{%
  \institution{Computing Product Line, Huawei}
  \city{Shanghai}
  \country{China}}

\author{Junsong Wang}
\email{junsongwang@huawei.com}
\orcid{0009-0006-5954-6958}
\affiliation{%
  \institution{Computing Product Line, Huawei}
  \city{Beijing}
  \country{China}}

\author{Jingwen Leng}
\email{leng-jw@sjtu.edu.cn}
\orcid{0000-0002-5660-5493}
\affiliation{%
  \institution{Shanghai Jiao Tong University}
  \city{Shanghai}
  \country{China}}
\affiliation{%
  \institution{Shanghai Qi Zhi Institute}
  \city{Shanghai}
  \country{China}}
\authornotemark[1]

\renewcommand{\shortauthors}{Weiming Hu et al.}

\begin{abstract}

Existing low-bit Microscaling (MX) formats, such as MXFP4, often suffer from substantial accuracy degradation due to the use of a shared scaling factor with the Power-of-Two format. 
In this work, we explore strategies that introduce minimal metadata to recover accuracy lost during quantization while maintaining high bit efficiency across a wide range of large language models.
We propose a complete algorithm-hardware co-design based on flexible metadata, featuring an online quantization with simple encoding.
To support the proposed method efficiently, we implement a lightweight hardware unit and integrate it into the accelerator. 
Evaluation results demonstrate that our method substantially narrows the accuracy gap, achieving on average a 70.63\% reduction in accuracy loss compared to MXFP4 and a 37.30\% reduction relative to the latest NVFP4 on LLM benchmarks. 
Furthermore, our design delivers up to 1.91$\times$ speedup and 1.75$\times$ energy savings over state-of-the-art accelerators.
Our code is available at \url{https://github.com/SJTU-ReArch-Group/M2XFP_ASPLOS26}.

\end{abstract}

\ccsdesc[500]{Computer systems organization~Single instruction, multiple data}
\ccsdesc[500]{Computer systems organization~Systolic arrays}
\ccsdesc[500]{Computer systems organization~Neural networks}

\keywords{Low-bit Quantization; Microscaling Data Formats; Hardware Acceleration}

\maketitle

\section{Introduction}\label{sec_intro}

Large language models (LLMs) have grown rapidly in scale and capability, with model size emerging as a primary driver of accuracy and generalization. 
\revise{State-of-the-art deployments now involve hundreds of billions of parameters, such as LLaMA-3.1~\cite{dubey2024llama3herdmodels}, which contains up to 405 billion parameters. }
Storing these models in standard BF16 precision alone requires terabytes of main memory, far exceeding the capacity of commodity accelerators. The resulting memory and compute demands place tremendous stress on both cloud-scale and device-level systems, motivating aggressive model compression techniques~\cite{lin2023awq,frantar2023gptq,ashkboos2024quarot,frantar2023sparsegpt,sun2024wanda,guan2024fractal,guan2022transkimmer,guo2020accelerating,guo2024accelerating}. 
Among these, low-bit quantization has emerged as a leading approach for reducing memory footprint, bandwidth consumption, and energy while preserving model quality, making it critical for the continued scaling of LLMs.

Recent advances in low-bit quantization have led to the adoption of Microscaling (MX) formats ~\cite{ocp_mx_specification}, which employ block-level shared scaling factors to enable fine-grained quantization. MX formats, such as MXFP4, have been widely adopted by industry and are natively supported in commercial accelerators, including NVIDIA’s B200~\cite{b200}, AMD’s MI300~\cite{amdmi300x}, and Microsoft’s Maia 100~\cite{maia100}. By exploiting shared exponents and streamlined dequantization, these formats deliver high throughput with minimal hardware overhead. However, accuracy degradation remains severe at 4-bit precision: the coarse resolution of power-of-two (E8M0) scaling misaligns with the local maximum, leading to significant rounding error, while more precise FP8 scaling (e.g., NVFP4) narrows dynamic range and requires additional rescaling\cite{jang2025blockdialect}. Other attempts, such as custom data types~\cite{guo2022ant,dettmers2023qlora,hu2025mant,chen2025bitmod,akshat2025microscopiq}, offer expressiveness but incur prohibitive hardware cost, especially for dynamic activations.

These limitations highlight a critical research gap: while scaling factor design has largely converged and data type innovations face scalability bottlenecks, the metadata axis remains relatively underexplored. Metadata, in principle, provides a flexible way to encode auxiliary precision or range information without altering the core data path, thereby enhancing quantization fidelity at low cost. Yet existing approaches remain fragmented. Outlier-oriented schemes (e.g., OliVe~\cite{guo2023olive}) improve accuracy in tensor-wise settings but break down in group-wise MX formats, while structural metadata (e.g., MicroScopiQ~\cite{ocp_mx_specification}) incurs excessive overhead, often exceeding 40 bits per block. Consequently, MX formats today either sacrifice accuracy for efficiency or burden hardware with metadata complexity, leaving a wide unexplored design space. This motivates the central question of our work: \textit{Can lightweight, principled metadata augmentation reconcile MX's efficiency with the accuracy demands of 4-bit LLM quantization?}
Our exploration therefore focuses on the metadata axis as the primary remaining degree of freedom to close the 4-bit accuracy gap.

To answer this, we propose \mymethod{} (Metadata-Augmented Microscaling Format), an algorithm–hardware co-design framework that systematically explores metadata allocation strategies. Our key insight is that metadata can serve as extra mantissa or exponent bits, enabling distinct trade-offs between precision refinement and range extension. Through a comprehensive design space exploration, we uncover a fundamental asymmetry: element-level metadata is most effective for dynamic activations, where lightweight, real-time encoding is essential, while subgroup-level metadata combined with scale search best serves static weights, where offline optimization is feasible. Building on this observation, we introduce a hybrid metadata scheme that applies element-level encoding to activations and subgroup-level encoding to weights. The resulting format improves bit efficiency with only 0.25 bits of metadata per element, delivering near-FP16 accuracy at effective 4.5-bit precision. We further design lightweight hardware support, integrated into systolic arrays with minimal extensions, that enables real-time metadata handling without disrupting the GEMM pipeline.

This paper makes the following contributions:

\begin{itemize}
    \revise{
    \item We introduce a taxonomy of MX design dimensions (scaling factor, data type, metadata). Unlike prior work that primarily varies scaling factors or base data types, we perform an EBW-guided design space exploration along the under-explored metadata axis, covering both element-level and subgroup-level metadata under fixed and adaptive shared scales.

    \item We identify an asymmetric behavior between weights and activations and propose \mymethod{}, a hybrid metadata-augmented MX format that uses element-level extra mantissa for activations and subgroup-level mantissa refinement with adaptive shared scales for weights.

    \item We design a hardware-efficient accelerator integration including a top-1 decode unit, an augmented FP4$\times$FP4 PE, and a streaming quantization engine, and show that \mymethod{} outperforms state-of-the-art MX accelerators in both accuracy and performance/energy at negligible area cost.
    }
\end{itemize}
\section{Background}\label{sec_bg}

\subsection{Model Quantization}

Quantization~\cite{lin2023awq,frantar2023gptq,ashkboos2024quarot,shao2024omniquant,tseng2024quipbetterllmquantization,dettmers2023qlora,lee2024tender,chen2025bitmod,guo2023olive,liu2025vqllm,lin2024duquant,guo2022squant} is a widely used technique for improving computational and memory efficiency by representing parameters with fewer bits.
The standard approach maps a full-precision tensor $\mathbf{X} \in \mathbb{R}^{n}$ to a low-precision grid via a single affine transformation.
Formally, for a target integer or floating point type with $k$-bit codes having dynamic range $[-Q_{\max}, Q_{\max}]$, each element is quantized as
\begin{equation}
    \tilde{x}_i = \text{round}\!\left(\frac{x_i}{s}\right), \quad 
    s = \frac{\max(|\mathbf{X}|)}{Q_{\max}},
\end{equation}
where $s$ is the per-tensor scaling factor and $\tilde{x}_i$ is stored as a $k$-bit integer or floating-point value.

Outliers are the primary cause of quantization error.
To mitigate the impact of outliers, recent studies~\cite{zhao2023atom,lin2023awq,dai2021vsquant,frantar2023gptq} reduce the quantization granularity from the tensor or channel level to finer groups. 
This approach, known as group-wise quantization, partitions the tensor's weights into small, fixed-size blocks (e.g., 64 or 128 values). 
Each block is quantized independently with its own unique scaling factor, effectively isolating the impact of any outliers within that small region.

\subsection{Microscaling Data Format}

\revise{Microscaling (MX) is an emerging low-bit data format widely supported by existing hardware~\cite{b200,maia100,amdmi300x}, and has been extensively studied and applied in recent works~\cite{mishra2025recipemxfp8,rtne_scale,akshat2025microscopiq,fang2025anda,lee2025mxplus,cuyckens2025mxrobotics,pmlr_sharify24a,lee2025amxfp4,koo2024opal,gil2025avantgarde,cook2025sixaccuratenvfp4quantization,zhang2025sageattention3,lo2023bsfp,khodamoradi2024errordiffusion,lo2023bucketgetter}.}
In this section, we introduce the standard MX format along with several of its variants.
Notably, the MX format naturally embraces group-wise quantization.

\begin{figure}[t] 
    \centering 
    \includegraphics[width=0.98\linewidth]{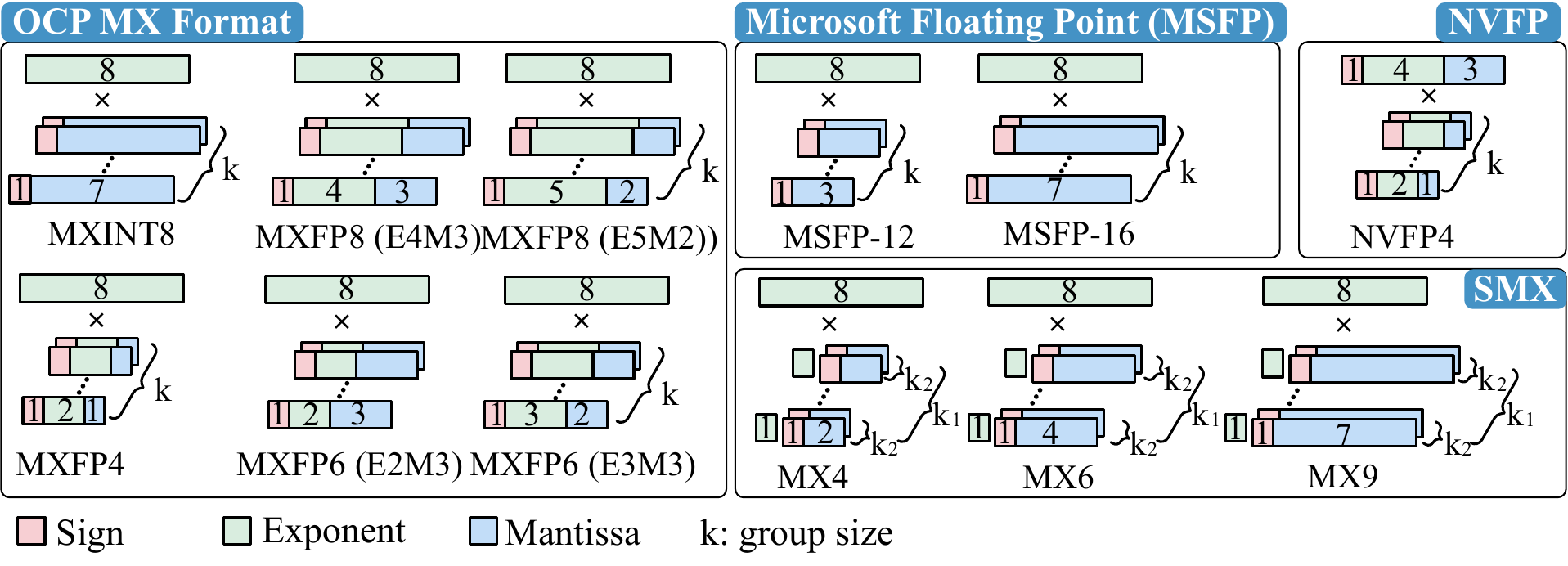}  
    \caption{Microscaling data format.}
    \label{fig:bg_mx}
\end{figure}

\textbf{Open Compute Project (OCP) Microscaling.}
Microscaling (MX) is a block floating-point format defined by the Open Compute Project (OCP)\cite{ocp_mx_specification}. 
As shown in \Fig{fig:bg_mx}, $k$ scalar elements share a common 8-bit scale factor. 
Unlike traditional group-wise quantization, where the scaling factor is FP16, the MX format restricts its scaling factor to the E8M0 format, a power-of-two representation with 8 exponent and 0 mantissa bits, making it particularly hardware-friendly for both quantization and dequantization processes.

\revise{
The shared scale is derived from the block maximum $x_{\max} = \max_i |V_i|$. 
Following the OCP specification~\cite{ocp_mx_specification}, the exponent of the shared scale is computed as $S = 2^{\lfloor \log_2(x_{\max} / P) \rfloor }$, where $P$ is the largest power-of-two representable in the target format (e.g., $P=
4$ for FP4). 
Recent works propose alternatives such as using $S = 2^{\lceil \log_2(x_{\max} / M) \rceil}$ instead to reduce clipping~\cite{mishra2025recipemxfp8}, where $M$ is the maximum representable value (e.g., $M=6$ for FP4), or incorporating rounding strategies like $S = 2^{\lfloor \log_2(\text{Round}(x_{\max}) / P) \rfloor}$~\cite{rtne_scale} to reduce systematic bias in scale selection. 
In this paper, we adopt the OCP-compliant floor-based method, and we will compare different scale calculations later.
}

The MX format quantization process can be simplified as a shift-and-rounding operation. 
For each element, its exponent is reduced by the shared exponent, which corresponds to a shift operation. 
Subsequently, the mantissa is rounded. 
The MX format dequantization process is seamlessly integrated into the General Matrix Multiplication (GEMM) operation in the latest modern GPUs~\cite{b200}.
Compared to conventional group-wise quantization, the dequantization process in MX format is more efficient and hardware-friendly.

\textbf{Variants of the Microscaling Format.}
Several variants of the MX format exist, all sharing a common feature: a shared scaling factor~\cite{NEURIPS2020_bfp,rouhani2023smx}.
The concept of block floating-point (BFP), also known as Microsoft Floating Point (MSFP)~\cite{NEURIPS2020_bfp}, was introduced by the Brain Project.
\Fig{fig:bg_mx} illustrates the MSFP-12 and MSFP-16 formats, where the numbers 12 and 16 refer to the combined bit widths of the scalar element and the shared scaling factor.

Microsoft and Meta have proposed Shared Microexponents (denoted as SMX in this paper)~\cite{rouhani2023smx}, which is a novel 2-level shared MX format.
The key distinction in this variant is that $k_2$ neighboring elements within share a 1-bit exponent, in addition to the 8-bit shared scaling factor by $k_1$ elements in a group.
Typically, $k_1$ is 16 and $k_2$ is 2.
The SMX family includes SMX4, SMX6, and SMX9, with differences in the mantissa bit width.
The number in the SMX format name corresponds to the combined bit width of the sign, shared exponent, and mantissa.

Recently, NVIDIA introduced NVFP, replacing the E8M0 scaling factor with the FP8 (E4M3) scaling factor. 
While FP8 scaling is more precise than E8M0, it has a reduced range, as the 4-bit exponent cannot cover the range of FP16. 
To compensate for this reduced range, NVIDIA proposes a tensor-level scaling factor to adjust the original tensor's distribution, making the FP8 scale factors more practical. 
This adjustment helps reduce quantization error by enhancing the precision of the scaling factor.
The 5th-generation tensor cores in NVIDIA's Blackwell architecture~\cite{b200} support both MXFP4 and NVFP4. 

\textbf{Key Takeaway.} Model quantization has evolved from coarse tensor-level schemes to fine-grained block-level formats, with Microscaling (MX) becoming the de facto hardware standard. MX achieves high throughput by exploiting shared power-of-two scaling and streamlined dequantization, and it has been widely adopted in commercial accelerators. However, this very reliance on a single shared scaling factor per block becomes a critical accuracy bottleneck at 4-bit precision, especially for LLM workloads where outliers dominate local dynamic ranges. Existing MX variants, e.g., MSFP, SMX, and NVFP, partially alleviate this issue but remain constrained by the same structural limitation, leading to either bit inefficiency or insufficient fidelity.

This gap motivates a deeper investigation into the design space of MX quantization, particularly exploring whether lightweight metadata augmentation can bridge the trade-off between bit efficiency and model accuracy. The next section analyzes the root causes of quantization error in MX formats and categorizes recent architectural optimizations, laying the groundwork for our proposed \mymethod{} design.

\section{Motivation}\label{sec_moti}

In this section, we first analyze the root causes behind the significant quantization error observed in low-bit MX formats.
We then summarize recent architectural optimizations designed to improve quantization performance in \Sec{sec_hw_analysis}.
Based on this, we categorize several optimization settings and evaluate them across different LLMs to identify the most effective configuration for MXFP.

\begin{figure}[t] 
    \centering 
    \includegraphics[width=0.98\linewidth]{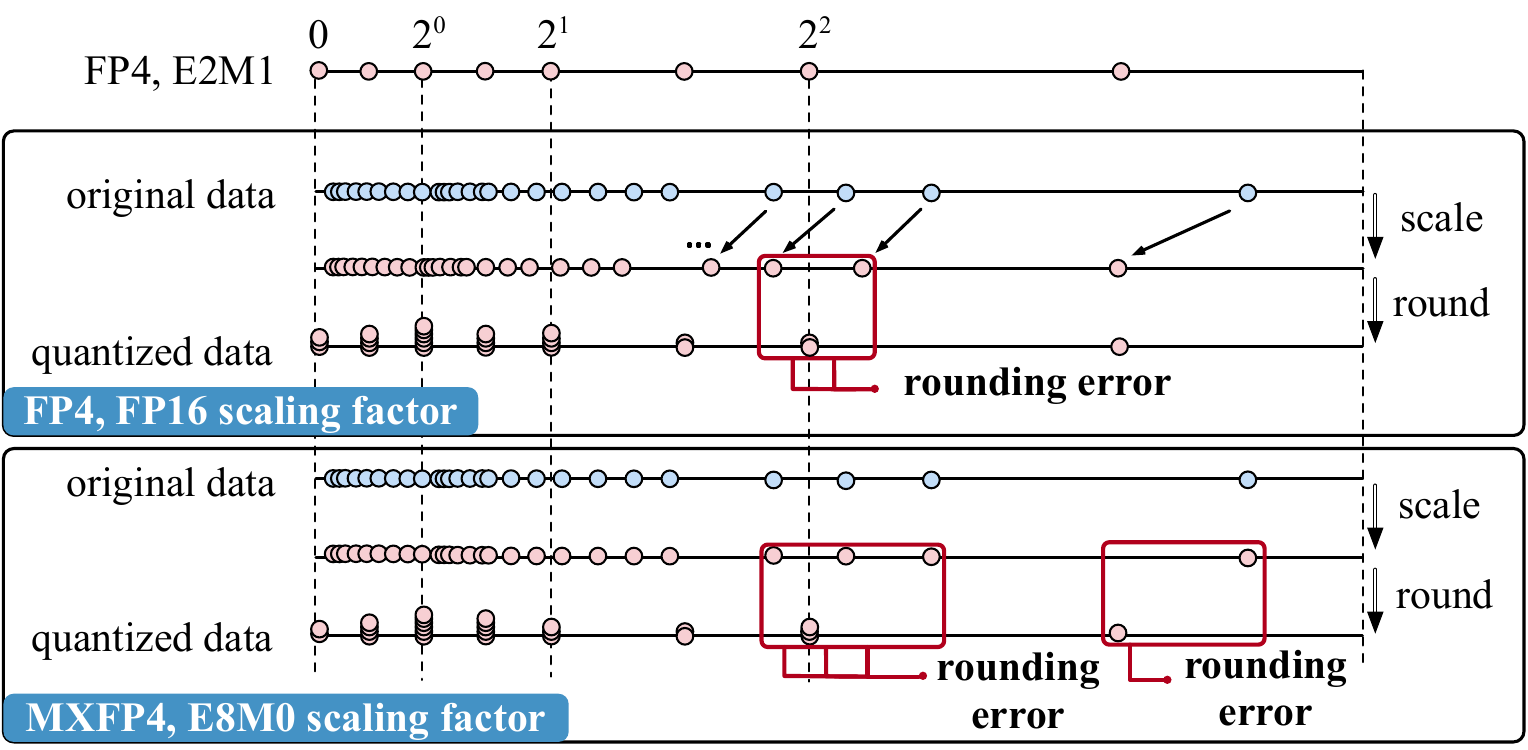}  
    \caption{FP4 quantization: A comparison of FP16 and E8M0 scaling factors.}
    \label{fig:moti_max_error}
\end{figure}

\subsection{Analysis of MX Quantization Error}
\label{sec:moti_max_error}

Low-bit MX formats suffer from significant accuracy degradation because their shared power-of-two scaling \textit{cannot} precisely align with block maximum~\cite{lee2025mxplus}. As illustrated in \Fig{fig:moti_max_error}, FP16-based scaling maps the maximum element of a group tightly to the FP4 maximum point, minimizing quantization error. In contrast, MX's E8M0 scaling only provides coarse power-of-two steps. When the group maximum falls between two exponent bins, the misalignment produces large rounding errors on the dominant value itself, which then propagates to the entire block.

\begin{figure}[t]
    \centering
    \begin{minipage}[t]{0.48\columnwidth}
      \centering
      \includegraphics[width=\columnwidth]{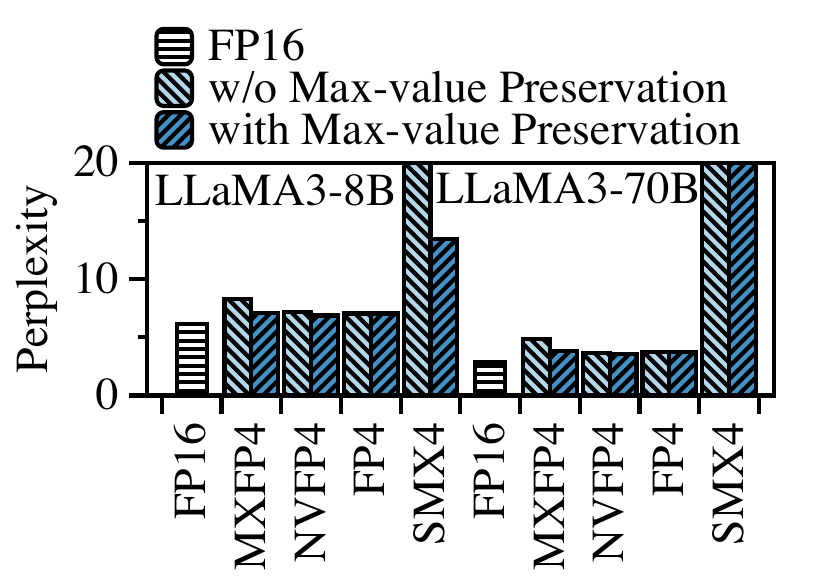}
      \caption{Perplexity of 4-bit quantization on LLaMA3, retaining the group-wise maximum in FP16 significantly enhances MXFP4.}\label{fig:moti_max} 
    \end{minipage}
    \hspace{2pt}
    \begin{minipage}[t]{0.48\columnwidth}
      \centering
      \includegraphics[width=\columnwidth]{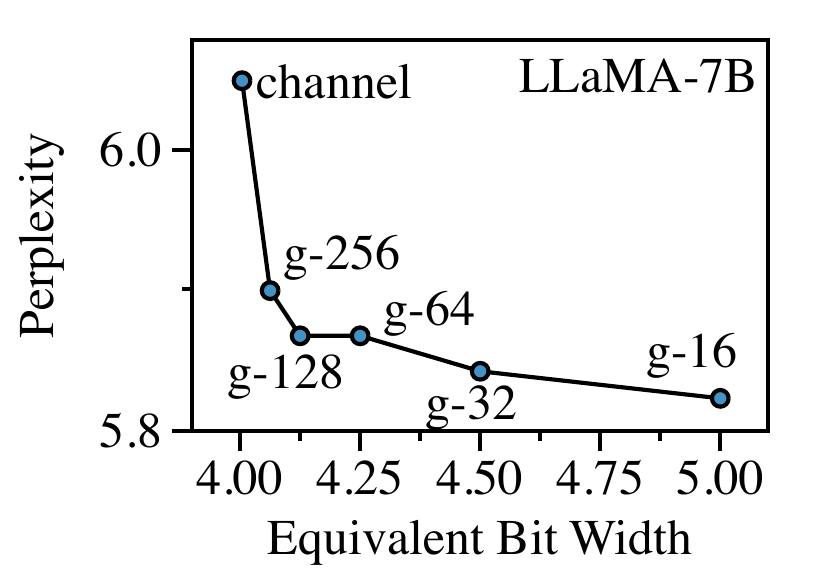}
      \caption{Perplexity decreases with increasing equivalent bit width (EBW), but the improvement diminishes beyond g-32 despite larger bit wdiths. }\label{fig:moti_ppl_ebw} 
    \end{minipage}
\end{figure}

We empirically validate this phenomenon by quantizing several LLMs with FP4, MXFP4, NVFP4, and SMX4, as shown in \Fig{fig:moti_max}. Both MXFP4 and SMX4 exhibit pronounced perplexity degradation. SMX4 performs especially poorly due to the additional shared 1-bit exponent among neighboring elements, which amplifies errors when their magnitudes differ. Crucially, we find that simply preserving the maximum element of the block in FP16 precision drastically reduces MXFP4's perplexity, nearly matching FP4 and NVFP4. This experiment confirms that the mishandling of block maximum is the primary weakness of MX quantization.

\begin{table*}[t]
    \centering
    \small
    \renewcommand{\arraystretch}{1.2}
    \caption[]{The features of DNN accelerators across different scaling factors, data types, and metadata designs are summarized. In the Scaling Factor column, `Granularity' refers to the quantization granularity. 
    In the Data Type column, a dash (`-') indicates that the architecture supports only a single data type. 
    }
  
    \resizebox{1.95\columnwidth}{!}{
      \begin{tabular}{c|cc|cc|cc}
        \Xhline{1.2pt}
        \multirow{2}{*}{Architecture} & \multicolumn{2}{c|}{Scaling Factor} & \multicolumn{2}{c|}{Data Type} & \multicolumn{2}{c}{Metadata}\\ \cline{2-7}
        & Granularity & Format & Granularity & Format  & Granularity & Content \\
        \Xhline{1.2pt}
        OliVe~\cite{guo2023olive} & Tensor/Channel & FP16 & Tensor/Channel & INT4, Flint4 & Pair & Outlier-victim pair  \\ 
        ANT~\cite{guo2022ant} & Tensor/Channel & FP16 & Tensor/Channel & INT4, Flint4, PoT4 & Tensor/Channel & 2-bit index   \\ 
        Tender~\cite{lee2024tender} & Channel & FP16 & - & INT4 & Channel & 12-bit index data \\ 
        MANT~\cite{hu2025mant} & Group-64 & FP16 & Group-64 & 16 data types & Group-64 & 8-bit coefficient $a$ \\ 
        BitMod~\cite{chen2025bitmod} & Group-128 & FP16 & Group-128 & FP4+special value & Group-128 & 2-bit index  \\ 
        \hline
        MXFP~\cite{ocp_mx_specification} & Group-32 & E8M0 & - & FP4 & - & -  \\ 
        SMX~\cite{rouhani2023smx} & Group-16 & E8M0 & - & \revise{INT3 (SMX4)} & Pair & 1-bit exponent \\ 
        NVFP~\cite{b200} & Group-16 & FP8 (E4M3) & - & FP4 & - & - \\ 

        \multirow{2}{*}{MicroScopiQ~\cite{akshat2025microscopiq}}  & \multirow{2}{*}{Group-128}  & \multirow{2}{*}{E8M0}  & \multirow{2}{*}{Group-128} & \multirow{2}{*}{FP4+INT4} & \multirow{2}{*}{Block} &  24-bit permutation list, 16-bit identifier,  \\ 
        &&&&&& and 8-bit MXScale, depends on $\mu\text{block}$ \\
        BBAL~\cite{han2025bbal} & Group-32 & E5M0 & - & INT3 & Element & 1-bit flag  \\ 
        BlockDialect~\cite{jang2025blockdialect} & Group-32 & E5M0 & Group-32 & 16 dialects & Group-32 & 4-bit index \\ 
        \revise{MX+~\cite{lee2025mxplus}} & Group-32 & E8M0 & Group-32 & FP4 & Group-32 & 5-bit index and 3-bit reserved \\ 
        \Xhline{1.2pt}
    \end{tabular}
    }
    \label{moti_arch_summary}
  \end{table*}

\subsection{A Taxonomy of Quantization Design Dimensions}
\label{sec_hw_analysis}

To further identify promising solutions, we decompose recent architectural innovations into three design dimensions: the scaling factor, the data type, and metadata.

\subsubsection{Scaling Factor: Converging Toward Group-Level E8M0/FP8}
The scaling factor determines how local dynamic ranges are represented. Early schemes adopted coarse per-tensor or per-channel scaling~\cite{frantar2023gptq,dettmers2023qlora}, which are simple but highly sensitive to outliers. Subsequent designs refined granularity to group-level scaling, partitioning tensors into blocks (e.g., 32 or 64 elements), each with its own shared scale. This approach, now embodied in OCP's MX specification ~\cite{ocp_mx_specification}, isolates local outliers and has become the industry standard. In terms of numerical format, the field has largely converged on two options: (1) \textbf{E8M0} (power-of-two scaling): extremely hardware-friendly due to its shift-only implementation, but coarse resolution leads to misalignment with block maximum (Sec. 3.1). (2) \textbf{FP8} (E4M3): higher precision, as adopted in NVIDIA Blackwell~\cite{b200}, but with limited exponent range, requiring an additional tensor-level rescale for stability.

As illustrated in \Fig{fig:moti_ppl_ebw}, our experiments show diminishing returns when simply reducing group size (e.g., from 32 to 16), while equivalent bit width (defined as per-element bits plus amortized scale bits) rises noticeably due to more scales per tensor, and accuracy gains quickly plateau. 

\revise{
In addition to scaling factor granularity, the data type of the scaling factor has also been explored in recent years to reduce storage overhead. 
\Tbl{moti_arch_summary} summarizes recent designs adopting E8M0 or FP8 formats. 
Overall, recent hardware and system designs converge on E8M0 and FP8 for scaling factors, suggesting that this dimension offers little room for breakthrough improvements.
}

\subsubsection{Data Type: Expressive but Hardware-Prohibitive}
Another line of work seeks to redesign the base data type to better match tensor distributions. This has led to a rich design space of specialized numerical formats, including custom types like Flint in ANT~\cite{guo2022ant}, non-uniform types in M-ANT~\cite{hu2025mant}, and the selectable `dialects' in BlockDialect~\cite{jang2025blockdialect}. 
While these methods provide strong representational flexibility, they face two fundamental limitations: (1) Low efficiency for dynamic tensors. Most designs target weights that are static and can afford offline type selection. Applying them to activations, which are generated dynamically during inference, requires costly runtime decisions. (2) Decoder complexity. Supporting multiple custom data types demands numerous decoders and format converters in hardware, significantly inflating area, latency, and energy. 

Thus, although novel data types are intellectually appealing, they pose significant challenges for deployment in low-latency, high-throughput accelerators, particularly for activation quantization, where runtime overhead is prohibitive.

\subsubsection{Metadata: A Flexible Yet Underutilized Design Axis}

Beyond scaling factors and data types, recent accelerators have begun exploring metadata~\cite{guo2023olive,akshat2025microscopiq,lee2024tender}, which apply small auxiliary bits that encode side information. Metadata can enhance accuracy without fundamentally altering the base data path, making it a lightweight yet versatile design axis. We identify three representative roles: (1) \textit{Critical-value precision allocation.} Approaches such as OliVe~\cite{guo2023olive} use “outlier-victim pairs” to assign extra bits to extreme values, while MicroScopiQ~\cite{akshat2025microscopiq} allocates different bit-widths to inlier and outlier blocks. (2) \textit{Range refinement.} SMX~\cite{rouhani2023smx} attaches a 1-bit secondary exponent to value pairs, and BBAL~\cite{han2025bbal} uses a 1-bit flag to shift exponents, both aiming to expand local dynamic range. (3) \textit{Format or structure control.} ANT~\cite{guo2022ant} and BlockDialect~\cite{jang2025blockdialect} employ metadata as indices for selecting numerical types, while Tender~\cite{lee2024tender} uses metadata to store indices to hint extra operations.

Despite their promise, existing metadata schemes remain fragmented and bit-inefficient. First, many focus on a single error source (e.g., outliers) but fail to address systemic quantization loss from block maximum. Second, others improve accuracy but at excessive control overhead (e.g., MicroScopiQ introduces 40+ bits of structural metadata per block). Last but not least, most works lack a principled framework for where and how to allocate metadata, especially for activations where both latency and hardware overhead are critical.

\subsection{Takeaway: Metadata as the Key Lever}

The analysis above highlights that scaling factor design has already converged, and data-type innovations face prohibitive hardware overheads for dynamic tensors. In contrast, metadata offers a flexible and underexplored design axis. Our findings indicate that properly allocating a small number of metadata bits (e.g., to preserve or enhance critical elements) can directly target the dominant error source in MX quantization. \textit{In summary, metadata remains the most underutilized yet most promising lever to close the gap between MX’s hardware efficiency and the accuracy demands of LLM quantization.}

\section{\mymethod{} Analysis and Design}
\label{sec_design}

In this section, we present the analysis and design of \mymethod{}, driven by an extensive encoding design space exploration (DSE) of metadata strategies. We first establish a unified framework for systematically reasoning about subgroup-level metadata allocation, then analyze Pareto trade-offs between accuracy and bit efficiency. Based on these insights, we derive a hybrid design tailored to the distinct characteristics of weights and activations, and finally detail the hardware-friendly quantization and encoding process.

\begin{figure}[t] 
    \centering 
    \includegraphics[width=0.98\linewidth]{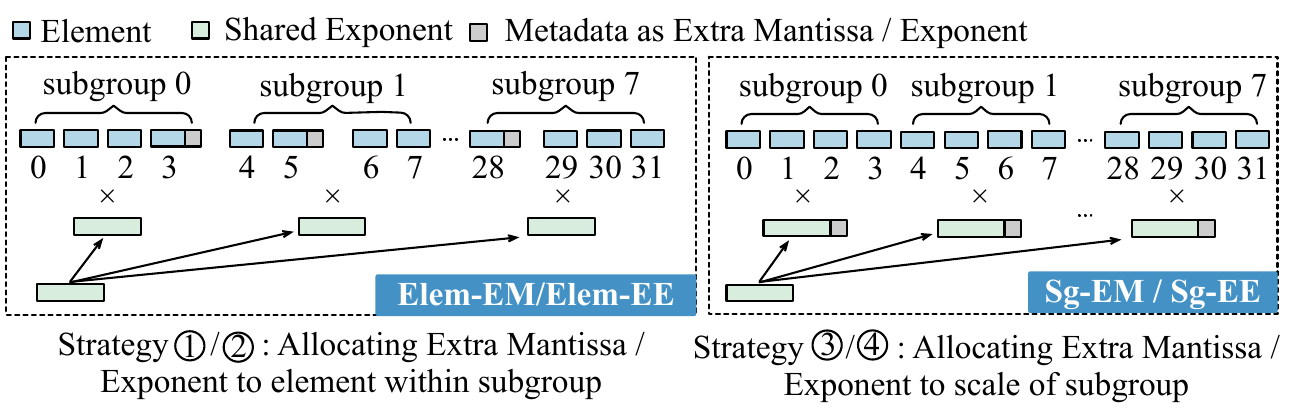}  
    \caption{
    MX format with subgroup-level metadata. The figure contrasts two strategies for allocating extra bits: (1) Elem-EM/EE extends individual elements within subgroups; (2) Sg-EM/EE augments the subgroup scale.}
    \label{fig:design_sgmx}
\end{figure}

\subsection{Framework for Design Space Exploration}
\label{sec:design_pre}

As discussed in \Sec{sec_hw_analysis}, metadata is the most flexible axis for improving MX quantization. To capture its full potential, we introduce a subgroup-centric framework that generalizes existing MX variants into a common design space. Specifically, a group of size \textit{k} is divided into \textit{N} contiguous subgroups, enabling localized metadata allocation. For instance, SMX can be interpreted as a group of 16 with subgroups of 2, each augmented by a 1-bit local exponent.

This unified abstraction allows us to organize metadata strategies along two orthogonal axes, as illustrated in \Fig{fig:design_sgmx}: (1) Precision vs. Range Enhancement: metadata can extend mantissa bits to refine precision or adjust exponent bits to expand dynamic range. (2) Element- vs. Subgroup-level Application: metadata can be applied to the most critical element within a subgroup or to the shared subgroup scale.

Focusing on hardware-feasible operations, we restrict metadata to mantissa or exponent augmentation, yielding four representative strategies:

\begin{itemize}
    \item Elem-EM (Element-level Extra Mantissa): use metadata as extra mantissa bits to a single element within each subgroup;
    \item Elem-EE (Element-level Extra Exponent): use metadata to provide an exponent offset to a single element;
    \item Sg-EM (Subgroup-level Extra Mantissa): enhance the precision of the subgroup scale factor, conceptually similar to NVFP4 but at finer granularity;
    \item Sg-EE (Subgroup-level Extra Exponent): encode a subgroup's exponent to improve local dynamic range, analogous to the SMX concept.
\end{itemize}

While metadata can be applied in many ways, its impact fundamentally depends on how it interacts with the group’s shared scaling factor (shared scale for short). Some strategies refine local precision without altering the global scale, whereas others allow metadata to reshape the scale itself. To clearly capture this distinction, we further define two allocation modes: (i) \textbf{fixed shared scale}: metadata locally refines elements or subgroups while leaving the group’s shared scale unchanged, acting as a lightweight precision or range enhancement, and (ii) \textbf{adaptive shared scale}: metadata also influences the choice of shared scale, enabling adaptive selection of scaling factors that minimize quantization error. 
\revise{
For instance, the fixed mode derives the shared scale strictly from the block maximum (i.e., $E=\lfloor \log_2(\text{amax}/P)\rfloor$), whereas the adaptive mode performs an MSE-based search over candidate exponents (e.g., $E, E\pm 1$) to jointly optimize the scale and metadata.
}

These two modes provide complementary perspectives. Fixed shared scale isolates the immediate effect of metadata bits with negligible complexity, while adaptive shared scale leverages metadata to globally rebalance quantization error. Taken together, they span the full metadata design space for MX quantization and establish a principled basis for the Pareto analysis in Sec. 4.2.

\begin{figure}[t] 
    \centering 
    \includegraphics[width=0.98\linewidth]{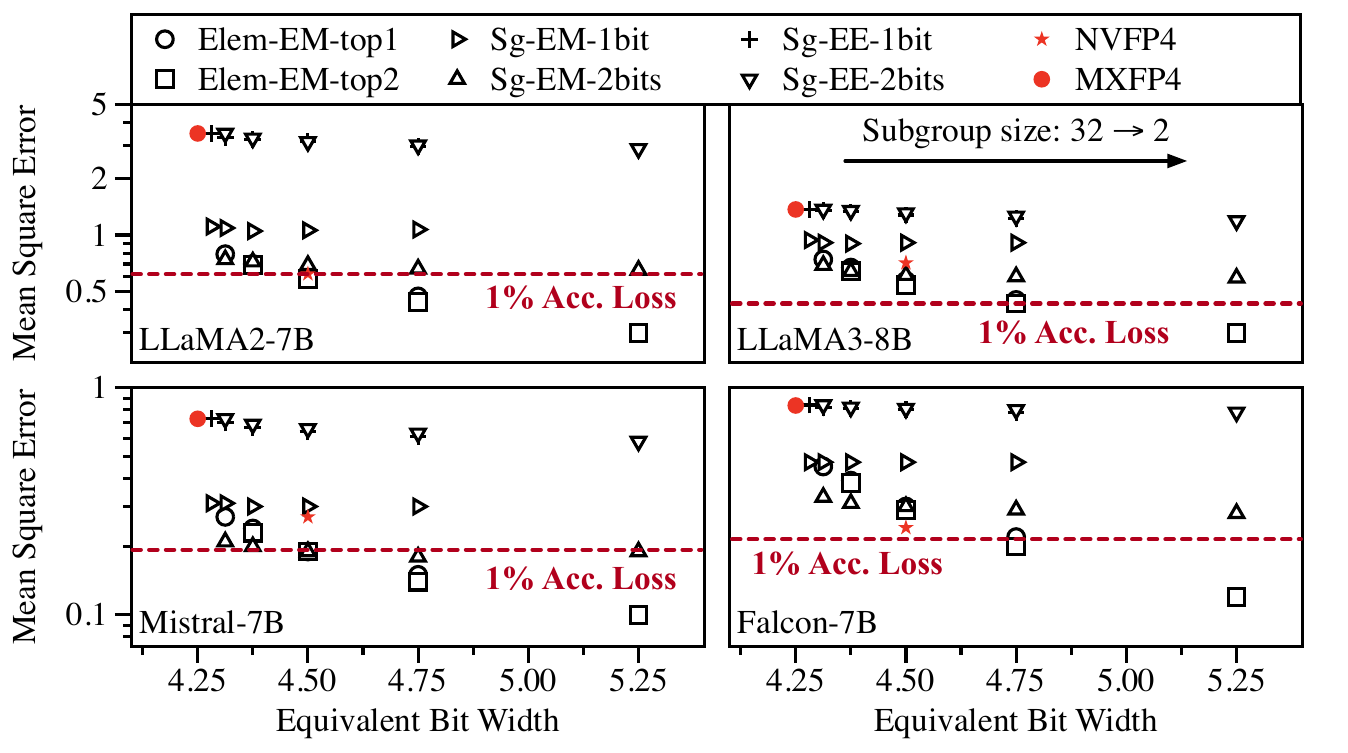}  
    \caption{Encoding design space exploration of Elem-EM, Sg-EM, and Sg-EE under fixed shared scale. Elem-EM achieves the lowest MSE at 4.5-4.75 EBW.}
    \label{fig:moti_dse}
\end{figure}

\subsection{Pareto-Optimal Analysis of Metadata Strategies}
\label{sec:pareto_analysis}

Building on the unified framework, we now evaluate the effectiveness of different metadata strategies under both fixed shared scale and adaptive shared scale modes. This analysis provides a principled way to characterize the trade-off between accuracy and bit efficiency, and identifies the Pareto-optimal configurations that best close the gap between MXFP4 efficiency and FP16 fidelity.

\subsubsection{Evaluation Method}
We quantify accuracy using mean squared error (MSE) relative to FP16. \revise{Specifically, the MSE is computed between the outputs of the quantized model, where both weights and activations are quantized, and those of the FP16 baseline, using the same input text. }We also normalize the storage cost using equivalent bit width (EBW), which incorporates element bits, shared scale, and metadata overhead, as shown in Eq.~\ref{eq:ebw}.
\begin{equation}
\label{eq:ebw}
\text{EBW} = \frac{(k \times B_{\text{elem}}) + B_{\text{meta}} + B_{\text{scale}}}{k} = B_{\text{elem}} + \frac{B_{\text{meta}} + B_{\text{scale}}}{k}
\end{equation}
Here, $k$ is the group size, $B_{\text{elem}}$ the base data bit-width (e.g., 4 for FP4), and $B_{\text{meta}}$ the metadata bits for the group. This metric allows fair comparisons across strategies by measuring the effective precision delivered per bit. 

To ensure fairness, the group size is fixed at 32, while subgroup size is varied to adjust EBW. For element-level strategies (Elem-EM), we assign 2 bits of mantissa metadata per element and evaluate both top-1 and top-2 allocations, but as \Sec{sec:moti_max_error} shows that exponent offsets cannot alleviate block-maximum errors, we omit Elem-EE. 
\revise{Here, top-1/top-2 denote the largest one or two values (by absolute magnitude) within each subgroup.}
For subgroup-level strategies (Sg-EM/EE), we allocate 1–2 bits of mantissa or exponent metadata to refine the shared scale. 
These configurations form the basis of the design space explored in Figs. 5–6 across LLaMA2-7B, LLaMA3-8B, Falcon-7B, and Mistral-7B. 
Experiments fix group size at 32 while varying subgroups to control EBW.

\subsubsection{Fixed Shared Scale Result (\Fig{fig:moti_dse}).}
Under a fixed shared scale, Elem-EM consistently dominates, achieving the lowest MSE in the 4.5–4.75 EBW range across all models. Top-1 and top-2 assignments yield nearly identical results, indicating that capturing only the maximum element per subgroup suffices. Sg-EM becomes competitive only at lower EBW ($\leq 4.375$), while Sg-EE shows negligible or marginal improvements over MXFP4 regardless of bit allocation. These results confirm that subgroup-level range expansion cannot address the dominant error source—block maximum misalignment. Notably, the red dashed line in \Fig{fig:moti_dse} marks the 1\% accuracy-loss threshold: Elem-EM reaches this target with ~4.6 bits on LLaMA2-7B, Falcon-7B, and Mistral-7B, and ~4.75 bits on LLaMA3-8B, whereas Sg-EM requires $\geq 5.25$ bits, and Sg-EE fails to meet the threshold entirely.
\begin{figure}[t] 
    \centering 
    \includegraphics[width=0.98\linewidth]{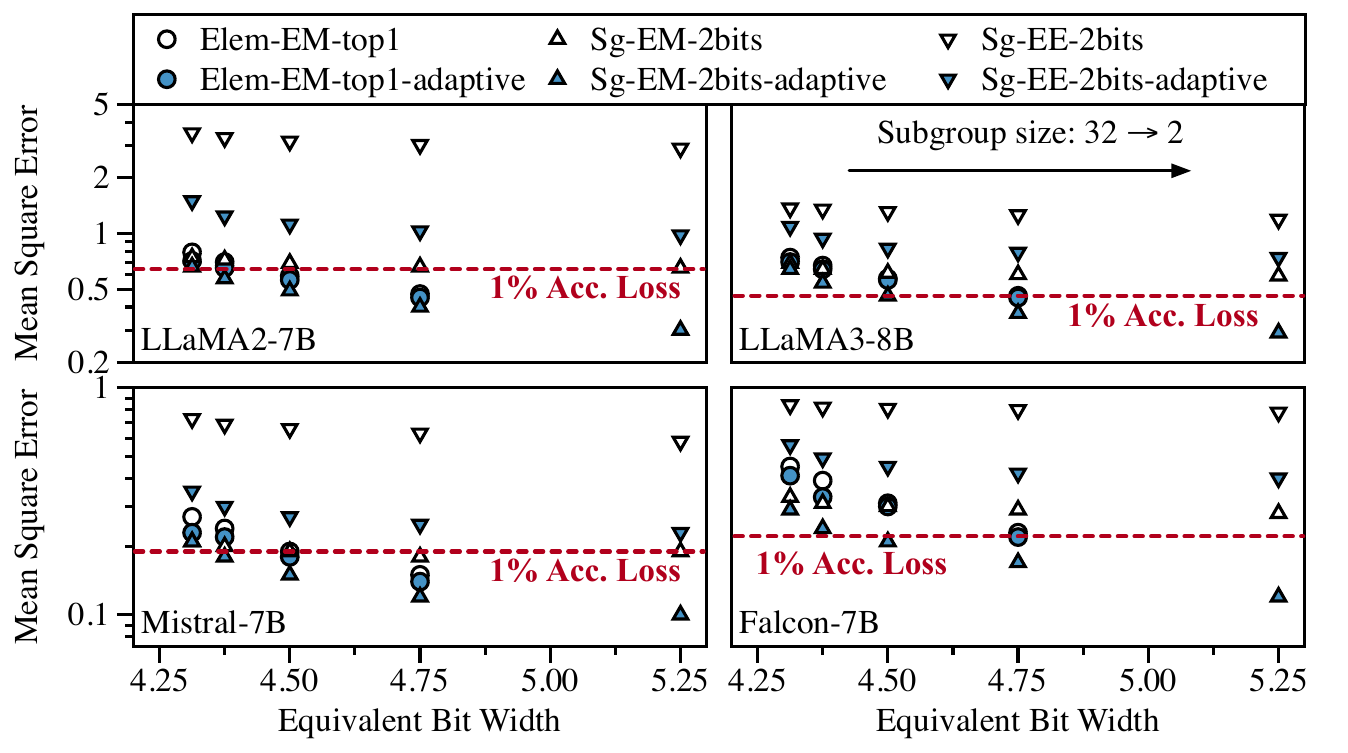}  
    \caption{Impact of adaptive shared scale on Elem-EM and Sg-EM. Optimizing rounding direction enables Sg-EM-search to outperform Elem-EM-search at 4.5-4.75 EBW. 
    }
    \label{fig:design_dse_search}
\end{figure}

\subsubsection{Adatpive Shared Scale Result (\Fig{fig:design_dse_search}).}
When adaptive shared scale is enabled, the Pareto frontier shifts. By adaptively selecting the shared scale in conjunction with metadata, Sg-EM-2bit surpasses Elem-EM in the critical 4.5 to 4.75 EBW region, achieving lower MSE with minimal overhead. Elem-EM still performs strongly, but no longer dominates. Sg-EE also benefits from adaptive shared scale, yet remains far less efficient than either Elem-EM or Sg-EM. Overall, the performance ranking becomes:
Sg-EM-adaptive > Elem-EM-adaptive > Elem-EM > Sg-EM > Sg-EE-adaptive > Sg-EE.

\subsubsection{Key Takeaway.} 
\revise{This Pareto analysis reveals a crucial asymmetry: element-level metadata is superior under a fixed shared scale due to its ability to capture dominant outliers without global adjustments, while subgroup-level metadata becomes preferable once an adaptive shared scale is incorporated.
\revise{
It leverages shared scale optimization to rebalance error across the block, which is quantitatively confirmed by the consistent MSE reduction for Sg-EM and Sg-EE in \Fig{fig:design_dse_search} (blue markers).
}
These complementary behaviors directly motivate the hybrid \mymethod{} design in \Sec{sec:m2xfp_design}, which assigns Sg-EM to static weights and Elem-EM to dynamic activations.}

\subsection{\mymethod{} Design: A Hybrid Strategy}
\label{sec:m2xfp_design}
The Pareto analysis highlights an important asymmetry: element-level metadata (Elem-EM) is the most effective under a fixed shared scale, while subgroup-level metadata (Sg-EM) becomes superior once an adaptive shared scale is incorporated. This observation naturally suggests that a single uniform strategy cannot simultaneously optimize for both weights and activations, which differ in their statistical properties and quantization requirements.

\textbf{Weights vs. Activations.} Weights are static and can be quantized offline, allowing sufficient time for adaptive optimization to identify the optimal subgroup-level refinement. In contrast, activations are generated dynamically during inference, where latency constraints demand lightweight, deterministic quantization. 
This constraint forces activations to adopt a bit-efficient strategy under the fixed shared scale mode.
As a result, weights benefit more from subgroup-level metadata with adaptive shared scale (Sg-EM-2bits-adaptive), while activations benefit from element-level metadata that directly preserves the most influential values (Elem-EM-top1).

\textbf{Hybrid Strategy.} \mymethod{} adopts a hybrid design that assigns: (1) Weights apply Sg-EM-2bit format, enabling fine-grained subgroup-scale refinement through offline adaptive optimization, thereby improving bit efficiency while maintaining fidelity. (2) Activations apply Elem-EM-top1 format, which captures outliers within each subgroup in real time with minimal routing overhead.

This division of labor leverages the strengths of both strategies while respecting the distinct hardware and workload constraints of weights and activations. Since Elem-EM-top1 and top2 show nearly identical accuracy, we adopt top1 for its simpler implementation and lower metadata routing complexity. As a result, \mymethod{} achieves near-FP16 accuracy at an effective precision of ~4.5 bits. 
This hybrid strategy establishes a balanced trade-off between hardware cost, quantization fidelity, and runtime efficiency, providing the foundation for hardware-friendly quantization and encoding.

\begin{figure}[t] 
    \centering 
    \includegraphics[width=0.95\linewidth]{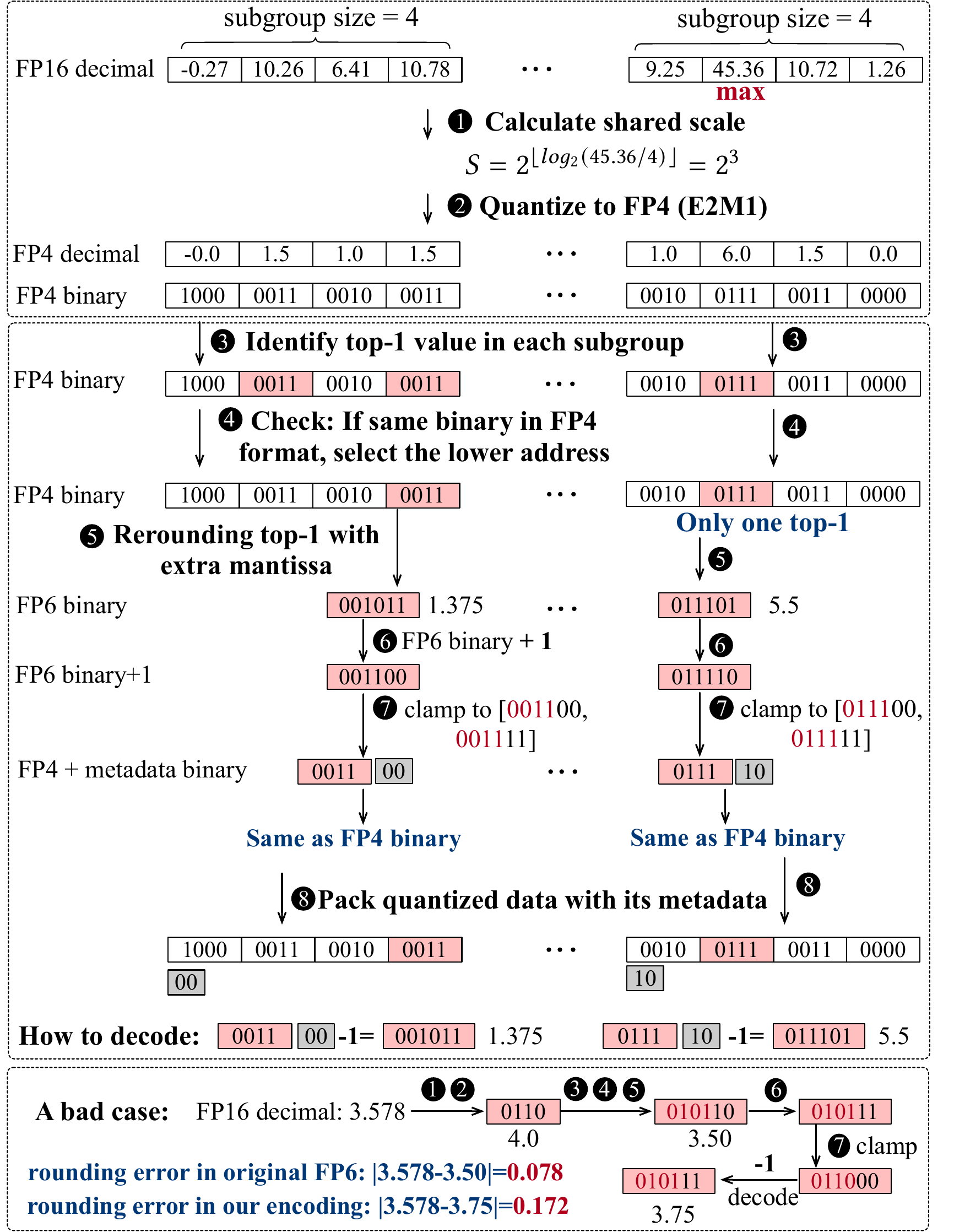}  
    \caption{Quantization process to \mymethod{} data format.}
    \label{fig:design_quantization}
\end{figure}

\begin{algorithm}
\caption{The \mymethod{} Quantization Process}
\label{alg:quantization_process}
\begin{algorithmic}[1]
    \State \textbf{Input:} High-precision data group $\mathbf{X}_{\text{FP16}}$ of size $k$.
    \State \textbf{Output:} Final MXFP4 $\mathbf{X}_{\text{FP4}}$ and metadata $\mathbf{X}_{\text{meta}}$.

    \Statex \textbf{\ding{182} Step 1: Calculate Shared Scale}
    \State $x_{max} \leftarrow \text{find maximum absolute value in } \mathbf{X}_{\text{FP16}}$
    \revise{\State $S \leftarrow 2^{\lfloor \log_2(\text{amax} / \text{FP4\_max\_pow2}) \rfloor} $}

    \Statex \textbf{\ding{183} Step 2: Quantize to FP4 (E2M1)}
    \State $\mathbf{X}_{\text{FP4}} \leftarrow \text{quantize\_to\_E2M1}(\mathbf{X}_{\text{FP16}}, S)$

    \Statex \textbf{For each subgroup:}
    \For{each subgroup $\mathbf{x}_{\text{FP4}}$ in $\mathbf{X}_{\text{FP4}}$}
        \Statex \textbf{\ding{184}\ding{185} Step 3 \& 4: Identify the top-1 in subgroup, resolving duplicates by selecting the lowest index}
        \State $\mathbf{x}_{\text{FP4\_abs}} \leftarrow abs(\mathbf{x}_{\text{FP4}})$
        \State $v_{max} \leftarrow \max(\mathbf{x}_{\text{FP4\_abs}})$ \Comment{Get max in subgroup}
        \State $C_{idx} \leftarrow \{ j \mid |\mathbf{x}_{\text{FP4\_abs}}[j]| = v_{max} \}$ \Comment{Get all candidate}
        \State $idx \leftarrow \min(C_{idx})$ \Comment{Select lowest index}
        
        \Statex \textbf{\ding{186} Step 5: Quantize top-1 to FP6(E2M3)}
        \State $x_{\text{orig}} \leftarrow \mathbf{X}_{\text{FP16}}[\text{idx}_{\text{top1}}]$ \Comment{Original value}
        \State $x_{\text{FP6}} \leftarrow \text{Quantize}(x_{\text{orig}}, \text{E2M3}, S)$
        \Statex \textbf{\ding{187} Step 6: Add bias for encoding}
        \State $\text{fp6\_bits} \leftarrow \text{FloatToBits}(|x_{\text{FP6}}|)$ \Comment{6-bit information}
        \State $\text{fp4\_bits} \leftarrow \text{FloatToBits}(|\mathbf{x}_{\text{FP4}}[\text{idx}_{\text{top1}}]|)$ \Comment{4-bit information}
        \State $\text{encoded} \leftarrow \text{fp6\_bits} + 1$ \Comment{Add bias in binary}
        \Statex \textbf{\ding{188} Step 7: Clamp to keep FP6 high 4 bits same as FP4}
        \State $\text{range\_min} \leftarrow \text{fp4\_bits} \underline{00}$\Comment{The minimum binary value with the same high 4 bits}
        \State $\text{range\_max} \leftarrow \text{fp4\_bits} \underline{11}$\Comment{The maximum binary value with the same high 4 bits}
        \State $\text{clamp} \leftarrow \text{Clamp}(\text{encoded}, \text{range\_min}, \text{range\_max})$
        \State $\mathbf{x}_{\text{meta}} \leftarrow \text{Get2BitsLow}(\text{clamp})$ \Comment{Extract 2-bit metadata}
        \Statex \textbf{\ding{189} Step 8: Pack quantized data with metadata}
        \State Append $\mathbf{x}_{\text{FP4}}$ to $\mathbf{X}_{\text{FP4}}$ and $\mathbf{x}_{\text{meta}}$ to $\mathbf{X}_{\text{meta}}$
    \EndFor

    \State \textbf{return} $\mathbf{X}_{\text{FP4}},\ \mathbf{X}_{\text{meta}}$
\end{algorithmic}
\end{algorithm}

\subsection{Quantization and Encoding Process}
\label{sec:quantization_process}

In this section, we introduce the quantization and encoding process for activation with Elem-EM and weight with Sg-EM.

\subsubsection{Activation Quantization with Elem-EM}
The online quantization process for \mymethod{} is designed to be efficient, as detailed in \Alg{alg:quantization_process} and illustrated in \Fig{fig:design_quantization}, with subgroup size 4 as an example.
For each incoming activation group, the group-level maximum absolute value is first determined to compute the shared scale factor (Step \ding{182}). 
All elements in the group are then quantized into a baseline 4-bit E2M1 representation (Step \ding{183}).

Given that the top1 maximum value within each subgroup must also be identified during decoding, we perform the selection in the 4-bit quantized format (FP4-E2M1) (Step \ding{184}).
In cases where multiple elements share the same maximum quantized value (i.e., different in FP16 but identical in FP4), \mymethod{} selects the element with the lowest memory address as the unique identification (Step \ding{185}). 
The original high-precision value of that identified top1 is then quantized to generate an FP6 value (Step \ding{186}).

\textbf{Encoding Strategy for FP6 Values.}
We identified a critical issue when directly replacing the FP4 value of the top1 element with its FP6 value: since the high 4 bits of the FP6 value are not necessarily identical to the original FP4 value, the top1 element may no longer remain the maximum after this replacement. To address this, we developed an improved encoding strategy.

Since quantization maps values to their nearest low-bit representation, a value quantized to a specific FP4 value $x$ has only five potential corresponding values when quantized to FP6. 
For example, if a value is quantized to 4 in FP4, it must fall within the range (3.5, 5]. 
Thus, it can only be quantized to one of 5 possible FP6 values: 3.5, 3.75, 4, 4.5, or 5.

Based on this observation, we can represent the FP6 value using a bias relative to the FP4 value. 
Centered at 4.0, the theoretical bias range is {-2, -1, 0, 1, 2}, corresponding to the FP6 candidates {3.5, 3.75, 4.0, 4.5, 5.0}.
However, for data alignment purposes, we clamp this bias to {-1, 0, 1, 2}, which introduces only minor rounding errors (in our example, rounding error occur only when a value is greater than 3.5 but less than 3.625). 
We give a case to indicate such a rounding error at the bottom of \Fig{fig:design_quantization}.
The additional rounding error introduced by our method has a negligible impact. Perplexity results show that the maximum deviation between results on common large language models with and without this rounding error is only 0.02, indicating a minimal effect on performance.

\textbf{Encoding Procedure.}
To implement this encoding, we first add a bias of 1 to the FP6 binary value, then clamp the result to ensure the high bits remain consistent with the original FP4 value. The lower 2 bits serve as metadata representing the additional mantissa precision.(Step \ding{187} \& \ding{188})

Finally, the definitive 4-bit values for the subgroup and the final 2-bit extra mantissa metadata are packed to form the final \mymethod{} representation (Step \ding{189}).
This packed data is organized in a hardware-friendly memory layout where the metadata for all subgroups is gathered into a single, contiguous block, followed by the 4-bit data elements. 


\subsubsection{Weight Quantization with Sg-EM}
Weight quantization is simpler than activation quantization. Each subgroup uses a 2-bit extra mantissa to refine the shared scale $S = 2^E$, giving candidates $\{1.0, 1.25, 1.5, 1.75\} \cdot S$. The search space for each subgroup is:
\begin{equation}
    \mathcal{S} = \{ (1+\tfrac{k}{4}) \cdot 2^{E} \mid k \in \{0,1,2,3\} \}
\end{equation}
When the adaptive shared scale is enabled, the whole group scale can be adjusted with a bias $b \in \{-1,0,1\}$ applied to the exponent. Notably, this bias requires no additional storage bits as it can be directly absorbed into the stored scale value. The optimal parameters are chosen via hierarchical MSE minimization:
\begin{equation}
    b^*, \{k_i^*\} = \arg\min_{b \in \{-1,0,1\}} \sum_{i \in \text{sg}} \left\| \hat{W}_{k_i^*,b} - W_i \right\|_2^2
\end{equation}
where $k_i^* = \arg\min_{k \in \{0,1,2,3\}} \left\| \hat{W}_{k,b} - W_i \right\|_2^2$, $W_i$ represents the original weights in subgroup $i$, and $\hat{W}_{k,b}$ denotes the weights quantized and dequantized using scale $(1+\tfrac{k}{4}) \cdot 2^{E+b}$.The optimization first finds the optimal mantissa refinement $k$ for each subgroup given a bias $b$, then selects the best group-level exponent bias $b$.

Concretely, for an 8-element subgroup, exploring all exponent and mantissa combinations requires evaluating up to 12 candidate scales, amounting to roughly $3 \times 8 \times 12 = 288$ FLOPs plus comparisons per subgroup. 
This overhead is acceptable for offline weight calibration but prohibitive for runtime activation quantization.

\section{Architecture}\label{sec_arch}
\begin{figure}[t] 
    \centering 
    \includegraphics[width=0.98\linewidth]{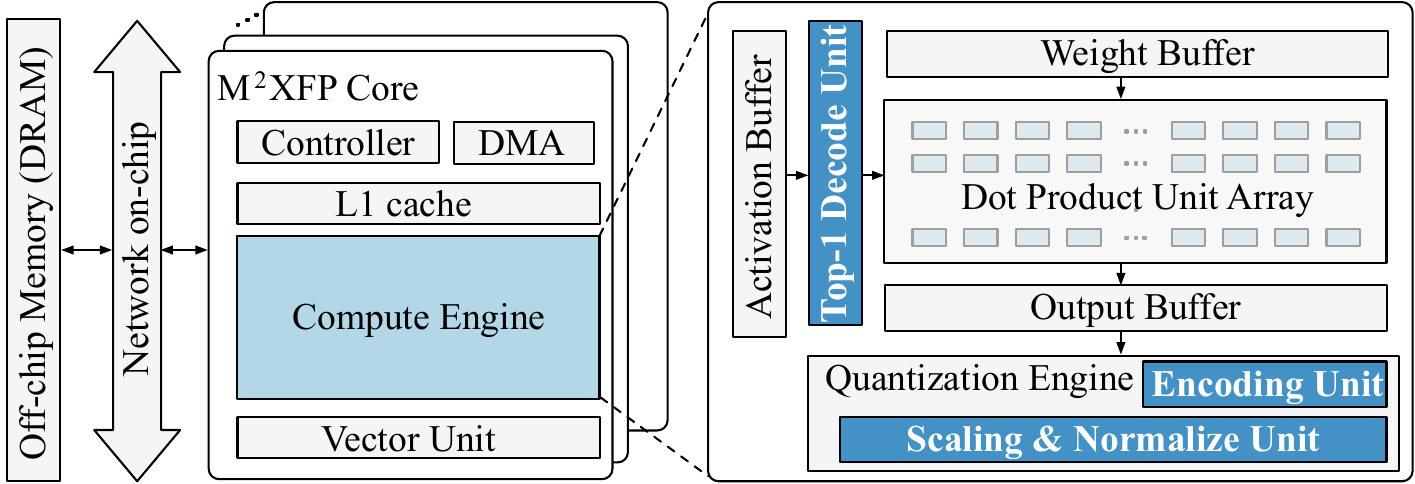}  
    \caption{Architecture Overview.}
    \label{fig:arch_overview}
\end{figure}
We now describe the architectural extensions required to efficiently support \mymethod{}. The design goal is to retain the high throughput and memory efficiency of systolic-array accelerators while introducing minimal logic to handle metadata and outlier refinement.\Fig{fig:arch_overview} illustrates the overall compute core, which integrates \mymethod{} into a conventional systolic array pipeline with lightweight modifications.

\subsection{Architecture Overview}

\begin{figure}[t] 
    \centering 
    \includegraphics[width=0.98\linewidth]{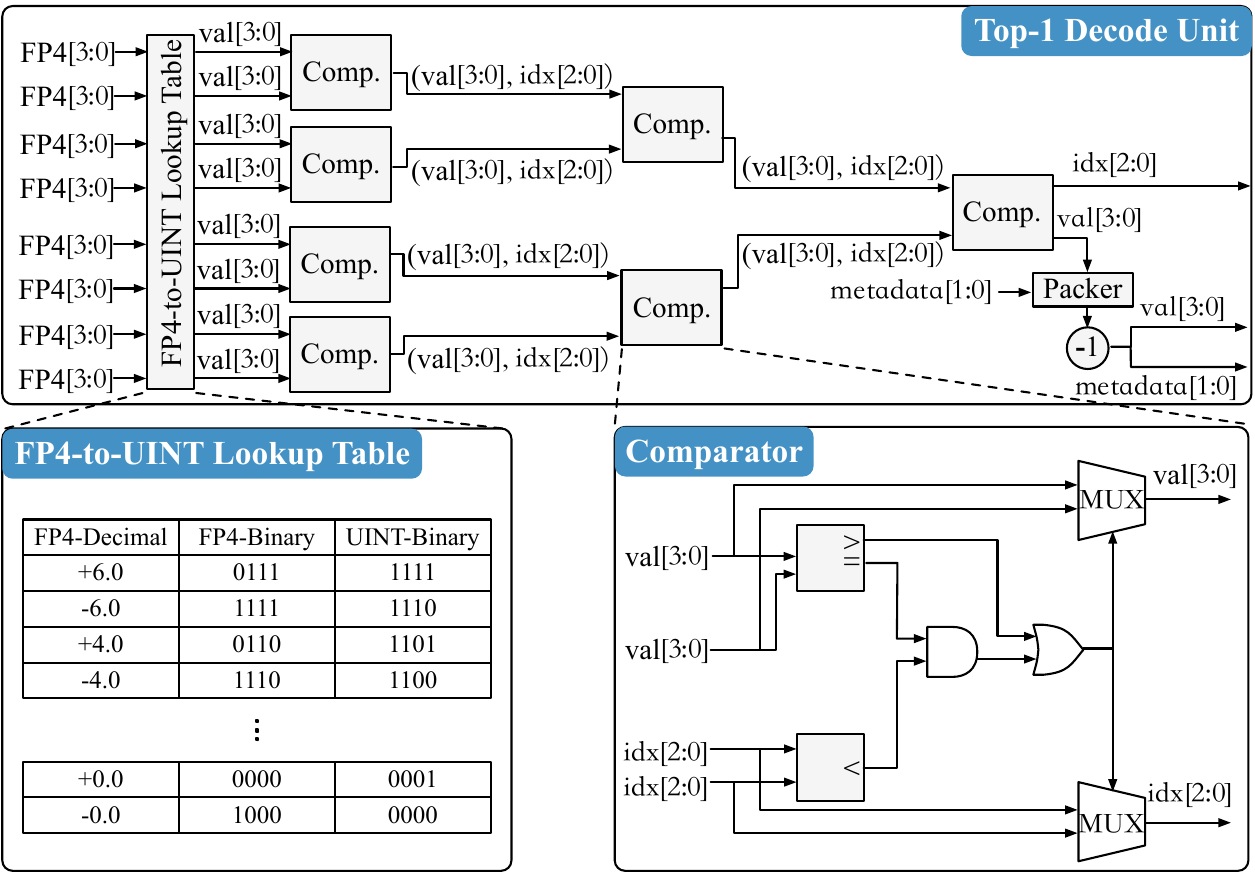}  
    \caption{Microarchitecture of the Top-1 Decode Unit, consisting of an FP4-to-UINT lookup table, a three-level comparator tree, and supporting logic.}
    \label{fig:arch_decoder}
\end{figure}

The baseline core includes standard components such as L1 cache, vector units, and on-chip buffers. \mymethod{} introduces three specialized units to support its hybrid quantization format:(1) \textbf{Top-1 Decode Unit} identifies the maximum element in each subgroup and forwards its metadata to the processing elements (PEs). (2) \textbf{Augmented PE Tile} executes FP4 × FP4 multiply–accumulate (MAC) operations while incorporating additional mantissa and subgroup-scale refinements. (3) \textbf{Quantization Engine} performs online Elem-EM quantization of activations following the procedure in \Sec{sec:quantization_process}. These units are strategically inserted between the input buffers and the systolic array to ensure seamless integration with existing GEMM pipelines.

\subsection{Memory Organization}

\mymethod{} preserves memory alignment by storing elements, scaling factors, and metadata in fixed-length fields.
Each group consists of three separately organized streams: a 128-bit block of packed 4-bit elements, an 8-bit shared scale, and 8-bit metadata.
These components are stored in contiguous memory regions, where elements in one continuous space, scale factors in another, and metadata in a third.
This separation not only guarantees alignment but also simplifies indexing and parallel access. Upon loading into on-chip buffers, a dispatch unit delivers the scale factor, metadata, and elements to the decode unit and PE array. 
This layout ensures that metadata handling introduces no fragmentation or misalignment overhead compared to baseline MXFP.

\subsection{Decode Unit}

The decode unit preprocesses input subgroups before computation. As shown in \Fig{fig:arch_decoder}, each FP4 element is first mapped to an unsigned integer using a compact 16-entry lookup table, enabling monotonic comparisons. A three-level comparator tree then identifies the unique top-1 element per subgroup. In case of ties, the lowest index is chosen, ensuring deterministic results.

The selected index and metadata are then packed and forwarded to the PE array. This logic is lightweight, comprising only a small LUT, comparators, and a multiplexer, yet crucial for enabling the element-level refinement in \mymethod{} without disrupting the systolic pipeline.

\begin{figure}[t] 
    \centering 
    \includegraphics[width=0.95\linewidth]{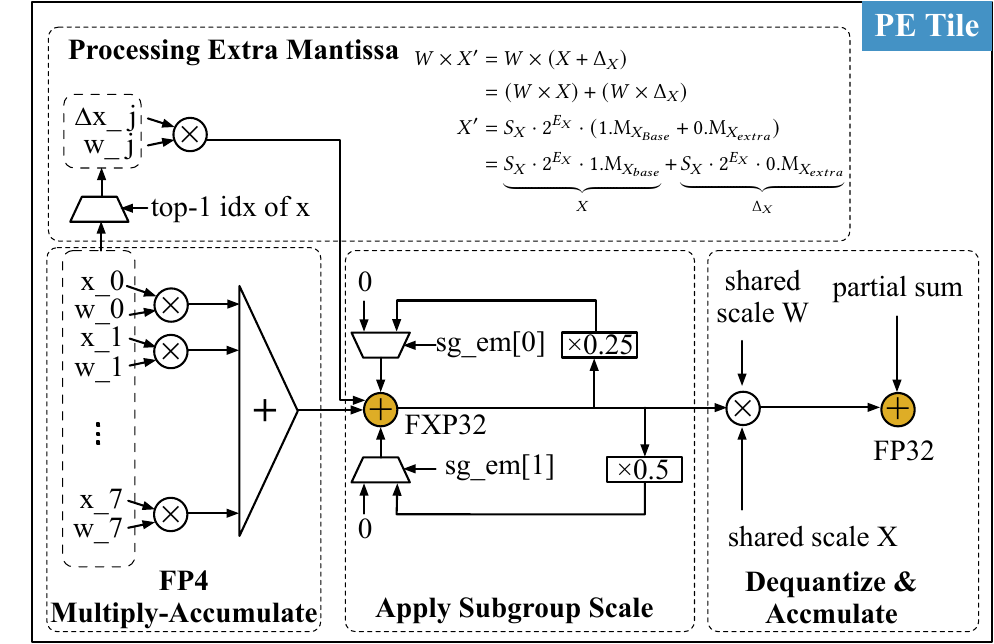}  
    \caption{Microarchitecture of the processing element.}
    \label{fig:arch_pe_array}
\end{figure}
\subsection{\mymethod{} Processing Element}

The PE tile is the key microarchitectural extension of \mymethod{}, as is shown in \Fig{fig:arch_pe_array}. Each PE integrates a baseline FP4-FP4 MAC datapath, augmented with logic for metadata refinement.

\textbf{Baseline FP4 MAC.}
The main datapath implements a conventional FP4 $\times$ FP4 MAC pipeline that processes the majority of subgroup elements. Eight parallel multipliers and an adder tree process subgroup elements using FP4 encodings, producing partial sums stored in a 32-bit fixed-point register. This provides sufficient dynamic range while facilitating efficient downstream scaling.

\textbf{Extra Mantissa Processing.}
To efficiently support elements with extended mantissa, we exploit the distributive property of multiplication. 
Given an element $X'$ with a 2-bit mantissa extension, it can be decomposed as $X' = X + \Delta X$, where $X$ is the FP4 baseline value and $\Delta X$ is a small correction term. 
Accordingly, the product expands to $W \times X' = W \times X + W \times \Delta X$. 
The term $W \times X$ is handled by the standard FP4 MAC pipeline, while the correction term $W \times \Delta X$ is computed in a lightweight auxiliary MAC unit and then accumulated with the baseline result. As illustrated in \Fig{fig:arch_pe_array}, the baseline component is represented as $S_X \cdot 2^{E_X} \cdot 1.\text{M}_{X_{base}}$, whereas $\Delta X$ corresponds to $S_X \cdot 2^{E_X} \cdot 0.\text{M}_{X_{meta}}$. The hidden bit of $\Delta X$ is set to zero, preserving compatibility with FP4 hardware and avoiding datapath disruption.

\textbf{Subgroup Scale Refinement.}
After intra-subgroup accumulation, each partial sum $P_i$ is adjusted by a subgroup-level scale determined by its 2-bit Subgroup Extra Mantissa ($sg_em$).
The $sg_em$ encodes fractional mantissa extensions corresponding to multipliers of $1.0$, $1.25$, $1.5$, and $1.75$ for codes \texttt{00}, \texttt{01}, \texttt{10}, and \texttt{11}, respectively.
These scaling factors can be efficiently realized using lightweight shift-and-add operations: $0.25P$ corresponds to a 2-bit right shift, $0.5P$ corresponds to a 1-bit right shift, and $0.75P$ is implemented by combining them ($0.5P + 0.25P$).
As the fixed-point datapath already provides sufficient range, no costly multipliers are needed, and the process incurs only minor hardware overhead.
The entire scaling procedure is illustrated in \Fig{fig:arch_pe_array}.

\begin{figure}[t] 
    \centering 
    \includegraphics[width=0.9\linewidth]{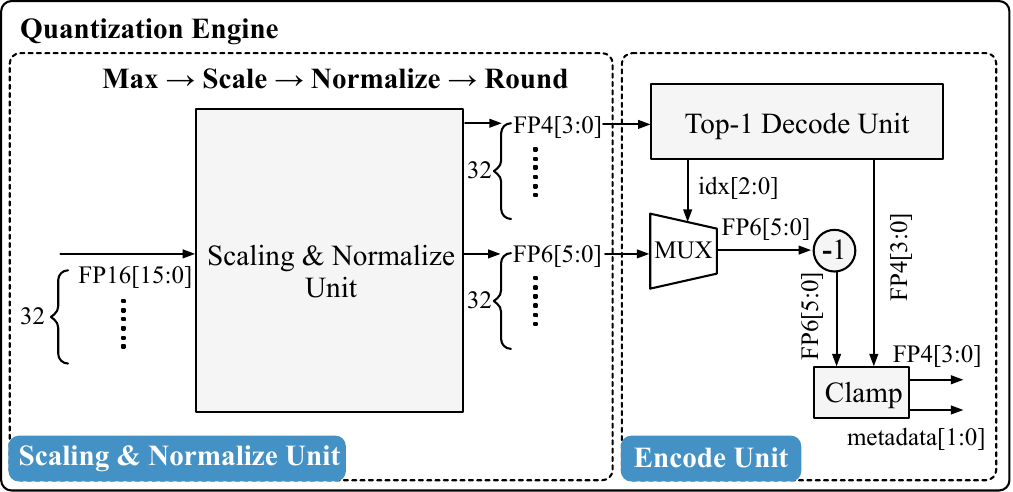}  
    \caption{The quantization engine that contains scaling \& normalization unit for quantization and encode unit to pack data to \mymethod{} format.}
    \label{fig:arch_quantization}
\end{figure}

\textbf{Dequantize and Accumulate.}
Finally, the scaled partial sums are dequantized into FP32 and accumulated across all subgroups before being written into the output buffer.
The final group output is obtained by summing all subgroup results and applying the shared scale.
For MX formats with an E8M0 shared scale, this dequantization is particularly lightweight: instead of full floating-point multiplications, it reduces to simple exponent alignment.

\subsection{Quantization Engine.}
As shown in \Fig{fig:arch_quantization}, the quantization engine is a two-stage pipeline responsible for online encoding of activations. The first stage computes the group-level scale and generates FP4/FP6 candidates; the second stage identifies the top-1 element per subgroup, applies the bias-clamp encoding, and packs the resulting FP4 data with 2-bit metadata. The entire process is deterministic and streaming-friendly, enabling real-time quantization without stalling the systolic array.

\section{Evaluation}\label{sec_eval}

\subsection{Experimental Setup}
\textbf{Models and Benchmarks.}
We evaluate our \mymethod{} across diverse workloads to demonstrate its generality. 
For large language models (LLMs), we test on LLaMA-2~\cite{touvron2023llama2} (7B), LLaMA-3~\cite{dubey2024llama3herdmodels} (8B, 70B), OPT~\cite{zhang2022opt} (6.7B), Mistral~\cite{jiang2023mistral7b} (7B), and Falcon~\cite{almazrouei2023falcon} (7B), covering 7B-70B parameters. 
LLM benchmarks include Wikitext v2 and common sense QA tasks such as Arc-challenge, Arc-easy, HellaSwag, PIQA, WinoGrande, and BoolQ~\cite{clark2018thinksolvedquestionanswering,clark2019boolq,Bisk2020piqa,2022winogrande,zellers2019hellaswag}. 
To further show robustness on reasoning, we evaluate reasoning-oriented models like DeepSeek-R1-Distill-Qwen~\cite{deepseekai2025deepseekr1} (1.5B, 7B) on AIME, MATH-500, GSM8K, GPQA-Diamond, and LiveCodeBench~\cite{aime_1983_2024,rein2024gpqa,jain2024livecodebench,lightman2023math500,cobbe2021trainingverifier_gsm8k}.

\textbf{Algorithm Implementation.}
We implement the \mymethod{} quantization framework in PyTorch~\cite{paszke2019pytorch}, enabling precise modeling of both \mymethod{} and baseline formats. 
Evaluation is conducted using lm-evaluation-harness~\cite{eval_harness}. 
Our MXFP4 baseline follows the OCP standard with group size 32; NVFP4 adopts group size 16; SMX also uses group size 16 with subgroup size 2. 
For \mymethod{}, we configure a shared E8M0 scaling factor with group size 32 and subgroup size 8.
\revise{This configuration is empirically validated in \Sec{sec:pareto_analysis} as a near-Pareto-optimal trade-off between granularity and overhead, while matching the group-size choices of existing MX-capable hardware~\cite{amdmi300x,b200}.}

\textbf{Algorithm Baselines.}
We evaluate \mymethod{} against MXFP4, SMX, and NVFP4, the quantization formats supported by existing hardware~\cite{maia100,b200,amdmi300x}. 
We further examine the benefits of enhancing NVFP4 with our proposed metadata augmentation.

\textbf{Accelerator Implementation.}
We extend the open-source, cycle-level simulator DNNWeaver~\cite{sharma2016high} to model our accelerator. 
The augmented components, decode unit, processing elements (PEs), and quantization engine, are implemented in Verilog and synthesized using Synopsys Design Compiler with the TSMC 28 nm standard cell library at 500 MHz, providing power and area estimates.
On-chip buffer power and area are modeled with CACTI v7~\cite{cacti2017}.

\begin{table}[t]
    \centering
    \small
    \renewcommand{\arraystretch}{1.0} 
    \caption[]{Zero-shot evaluation results on five benchmarks: Arc-e (Arc-Easy), Arc-c (Arc-Challenge), Hella. (HellaSwag), PiQA, and Wino. (Winogrande). Group / subgroup sizes — MXFP4: 32 / 32, SMX: 16 / 2, \mymethod{}: 32 / 8.}
    \resizebox{0.98\columnwidth}{!}{
    \begin{tabular}{l|cccccc|c}
        \Xhline{1.2pt}
        \textbf{Method} & Arc-e & Arc-c & Hella. & PiQA & Wino. & BoolQ & Avg.  \\
        \hline
        & \multicolumn{6}{c}{LLaMA2-7B } \\
        \hline
        FP16 &  74.58 & 46.25 & 75.99 & 79.11 & 69.06 & 77.71 & 70.45  \\
        SMX4 & 26.43   & 27.05 & 26.13 & 49.40 & 49.80 & 38.93 & 36.29 \\
        MXFP4 & 66.84 & 41.47 & 70.49 & 76.61 & 64.01 & 72.51 & 65.32 \\
        NVFP4 & 73.11 & \textbf{44.88} & 74.62 & \textbf{78.13} & 67.88 & 74.22 & 68.81 \\

        \mymethod{}& \textbf{73.32} & 44.37 & \textbf{74.64} & 77.58 & \textbf{68.27} & \textbf{76.97} & \textbf{69.19}  \\

        \hline
        & \multicolumn{6}{c}{LLaMA3-8B} \\
        \hline
        FP16 & 77.49 & 53.33 & 79.15 & 80.85 & 72.53 & 81.28 & 74.11  \\
        SMX4 & 25.00 & 27.13 & 26.03 & 50.18 & 48.86 & 40.67 & 36.31 \\
        MXFP4 & 71.42 & 46.08 & 73.53 & 77.48 & 68.19 & 72.84 & 68.26 \\
        NVFP4 & 72.98 & 48.55 & 76.08 & 78.40 & \textbf{72.14} & 75.96 & 70.69 \\

        \mymethod{}& \textbf{74.58} & \textbf{49.57} & \textbf{77.23} & \textbf{79.54} & 70.96 & \textbf{79.20} & \textbf{71.85}  \\

        \hline
        & \multicolumn{6}{c}{Mistral-7B-v0.3} \\
        \hline
        FP16 & 78.24 & 52.13 & 80.46 & 82.26 & 73.8 & 82.14 & 74.84  \\
        SMX4 & 26.39 & 27.22 & 25.69 & 49.18 & 49.33 & 40.06 & 36.31\\
        MXFP4 & 74.03 & 46.67 & 75.87 & 78.94 & 69.06 & 73.49 & 69.68 \\
        NVFP4 & 76.47 & 49.23 & 78.13 & \textbf{81.56} & 70.64 & 78.07 & 72.35 \\

        \mymethod{}& \textbf{76.64} & \textbf{50.85} & \textbf{79.76} & 80.74 & \textbf{71.27} & \textbf{82.45} & \textbf{73.62}  \\

        \Xhline{1.2pt}
    \end{tabular}
    }
  
    \label{eval:llm_acc}
  \end{table}

\textbf{Accelerator Baselines.}
We evaluate the performance and energy of \mymethod{} against representative accelerator baselines. 
Our primary comparison is with MicroScopiQ, a state-of-the-art (SOTA) MX-based accelerator that partitions weights into inlier and outlier blocks, applying hybrid MX quantization to weights and MXINT to activation. 
To a comprehensive evaluation, we adapt non-MX accelerators (ANT, M-ANT, OliVe) to support fine-grained MX quantization, denoted MX-ANT, MX-M-ANT, and MX-OliVe. 
We also include BlockDialect, an SOTA algorithm-architecture co-design approach, with perplexity in \Sec{sec:baseline_accelerator}. 

For fairness, all accelerators are configured with 32$\times$32 PEs supporting 4-bit multiplications, ensuring differences arise from architectural and algorithmic design.

\begin{table}[t]
    \centering
    \small
    \renewcommand{\arraystretch}{1.0} 
    \caption[]{Perplexity on the Wikitext dataset for \mymethod{} and baseline accelerators (lower is better).}
    \resizebox{0.98\columnwidth}{!}{
    \begin{tabular}{l|cccccc}
        \Xhline{1.2pt}
        \textbf{Method} & LLaMA2 & LLaMA3 & LLaMA3 & OPT & Mistral & Falcon    \\
         & 7B & 8B & 70B & 6.7B & 7B & 7B    \\
        \hline
         FP16 & 5.47 & 6.14 & 2.85 & 10.86 & 5.32 & 6.59\\
         MXFP4 & 7.15 & 8.30 & 4.84 & 19.21 & 6.56 & 7.59\\
        MX-ANT & 6.30 & 8.22 & 4.65 & 12.76 & 6.04 & 7.35  \\
        MX-M-ANT & 6.12 & 7.83 & 4.54 & 12.45 & 5.89 & 7.32  \\
        MX-OliVe & 7.46 & 11.33 & 6.84 & 36.80 & 6.77 & 8.40   \\
        MicroScopiQ & 6.24 & 8.33 & 4.75 & 12.65 & 6.00 & 7.45   \\
        BlockDialect & 5.84 & 7.05 & 3.76 & \textbf{11.31} & 5.65 & 6.94 \\
        \mymethod{}& \textbf{5.77} & \textbf{6.84} & \textbf{3.56} &  11.34 & \textbf{5.58} & \textbf{6.88}  \\
        \hline
        
        \Xhline{1.2pt}
    \end{tabular}
    }
  
    \label{eval:llm_ppl}
  \end{table}

\subsection{Large Language Model Evaluation}

\textbf{Compared to Existing Data Types.}
We first evaluate accuracy on LLMs, with results summarized in \Tbl{eval:llm_acc}, comparing \mymethod{} against several hardware-supported data types. 

SMX4 shows severe degradation, with average accuracy loss exceeding 30\% on 7B/8B models, making it impractical for 4-bit weight-activation quantization. 
MXFP4 is more stable, with an average loss of 5.38\% on 7B/8B. 

\mymethod{} consistently outperforms MXFP4 across all model scales. 
Specifically, on 7B/8B models, the average accuracy loss is reduced to 1.58\%, representing a 70.63\% improvement over MXFP4. 
Compared with NVFP4 at the same effective bit-width (4.5 bits), \mymethod{} also shows lower loss (1.58\% vs. 2.52\%), corresponding to an absolute accuracy gain of 0.94\% (37.30\% improvement). 
We note that NVFP4 achieves higher accuracy on certain tasks (e.g., WinoGrande on LLaMA3-8B), which demonstrates its effectiveness. 
However, when averaged across all benchmarks, \mymethod{} consistently delivers superior overall accuracy.
\revise{
Other data types also benefit from adaptive shared scale search, but these gains do not change the overall trends in \Tbl{eval:llm_acc}.
}

\textbf{Compared to Baseline Accelerators.}
\label{sec:baseline_accelerator}
We evaluate perplexity on Wikitext v2 across several LLMs, comparing \mymethod{} with MX-ANT, MX-M-ANT, MX-OliVe, MicroScopiQ, and BlockDialect, all under W4A4 quantization with group size 32 and an E8M0 shared scaling factor.

\mymethod{} achieves the lowest perplexity on all models except OPT-6.7B, where BlockDialect is better by only 0.03.
MX-ANT and MX-M-ANT improve over MXFP4 by adapting weight types, but extending to activations is limited by costly online search. 
BlockDialect addresses this with efficient real-time decision, yielding larger gains over MXFP4.
MX-OliVe, though effective tensor-wise, underperforms MXFP4 in group-wise due to its `outlier-victim' encoding that sacrifices neighbors. MicroScopiQ shows that such neighboring outliers frequently occur in LLMs, and adopts a block-level scheme for better balance. However, its reliance on naive MXINT activation quantization leads to suboptimal W4A4 perplexity.

\begin{table}[t]
    \centering
    \small
    \renewcommand{\arraystretch}{1.0} 
    \caption[]{Evaluation of reasoning tasks on DeepSeek-R1-Distill-Qwen: MXFP4 vs. \mymethod{}. }
    \resizebox{0.98\columnwidth}{!}{
    \begin{tabular}{l|ccccc|c}
        \Xhline{1.2pt}
        \textbf{Method} & AIME-90 & MATH-500 & GSM8K & GPQA & LiveCodeBench  & Avg.  \\
        \hline
        & \multicolumn{6}{c}{DeepSeek-R1-Distill-Qwen-1.5B } \\
        \hline
         FP16 & 21.11 & 85.4 & 84.76 & 36.36 & 17.54 & 49.03     \\
         MXFP4 & 7.78 & 66.6 & 69.37 & 31.82 & 8.96 & 36.91   \\
        \mymethod{}& 18.89 & 80.2 & 79.83 & 32.83 & 10.45 & 44.44  \\
        \hline
        & \multicolumn{6}{c}{DeepSeek-R1-Distill-Qwen-7B} \\
        \hline
         FP16 & 45.56 & 93.80 & 90.83 & 50.51 & 35.82 & 63.30     \\
         MXFP4 & 26.67 & 89.60 & 88.40 & 46.97 & 28.36 & 56.00   \\
        \mymethod{}& 40.00 & 93.80 & 90.83 & 52.02 & 32.40 & 61.81 \\
  
        \Xhline{1.2pt}
    \end{tabular}
    }
  
    \label{eval:llm_reasoning}
  \end{table}

\begin{table}[t]
    \small
    \renewcommand{\arraystretch}{1.25} 
    \caption{The area and power of core components and buffers for \mymethod{} using a 28nm process.}
  
    \resizebox{0.9\columnwidth}{!}{%
    \begin{tabular}{cccc}
    \Xhline{1.2pt}
    Component & Number & Area($mm^2$) & Power(mW) \\ \cline{2-4} 
    \Xhline{1.2pt}
    
    PE Tile (2140.12$\mu m^2$) & 128  & 0.2739 & 27.021  \\ 
    Top-1 Decode Unit (82.91$\mu m^2$) & 4 & 0.0003 & 0.064 \\ 
    Quantization Engine (2451.47$\mu m^2$) & 1 & 0.0024 & 0.663 \\ 
    Buffer (324KB)   & 1 & 0.7740 & 176.268 \\ 
    \hline
    Total &  & 1.051  & 204.02 \\ 
  
    \Xhline{1.2pt}
    \end{tabular}%
    }
    \label{tab:area}
  \end{table}
\textbf{Reasoning Tasks.}
We evaluate \mymethod{} on complex reasoning benchmarks using DeepSeek-R1-Distill-Qwen. 
Prior work~\cite{liu2025quantizationhurtsreasoning} shows that MXFP4 severely degrades reasoning ability, making LLMs nearly incapable of handling advanced math or coding tasks. 
Our results in \Tbl{eval:llm_reasoning} confirm this: MXFP4 causes a 12.12\% accuracy drop on DeepSeek-R1-Distill-Qwen-1.5B.
\mymethod{} can recover the average accuracy loss to 4.59\%.
Moreover, \mymethod{} scales robustly to 7B reasoning models, maintaining reliable performance across sizes.

\subsection{Performance, Area, and Energy}
\Tbl{tab:area} presents the component breakdown of \mymethod{}.
A $32 \times 32$ systolic array is modeled with four top-1 decode units, each handling eight 4-bit inputs. 
Together with the quantization engine, these account for only 0.26\% of area and 0.36\% of power overhead in all components, reflecting the low overhead of MX quantization in E8M0 format. 
\revise{
To quantify the hardware overhead across data formats, we synthesized MXFP4, NVFP4, and \mymethod{} PE tile using the same 28nm flow.
The resulting PE tile areas are 2057.6$\mu m^2$ (MXFP4), 2104.7$\mu m^2$ (NVFP4, +2.3\%), and 2140.1$\mu m^2$ (\mymethod{}, +4.0\%), showing that \mymethod{} remains in the same cost range as existing MX-based formats and introduces only modest additional area.
}
The design includes 324 KB of buffer: 144 KB each for activations and weights, plus 36 KB for outputs with scaling factors and metadata.
It is worth noting that the buffer size also incorporates storage for scaling factors and metadata.

\begin{figure}[t] 
    \centering 
    \includegraphics[width=0.98 \linewidth]{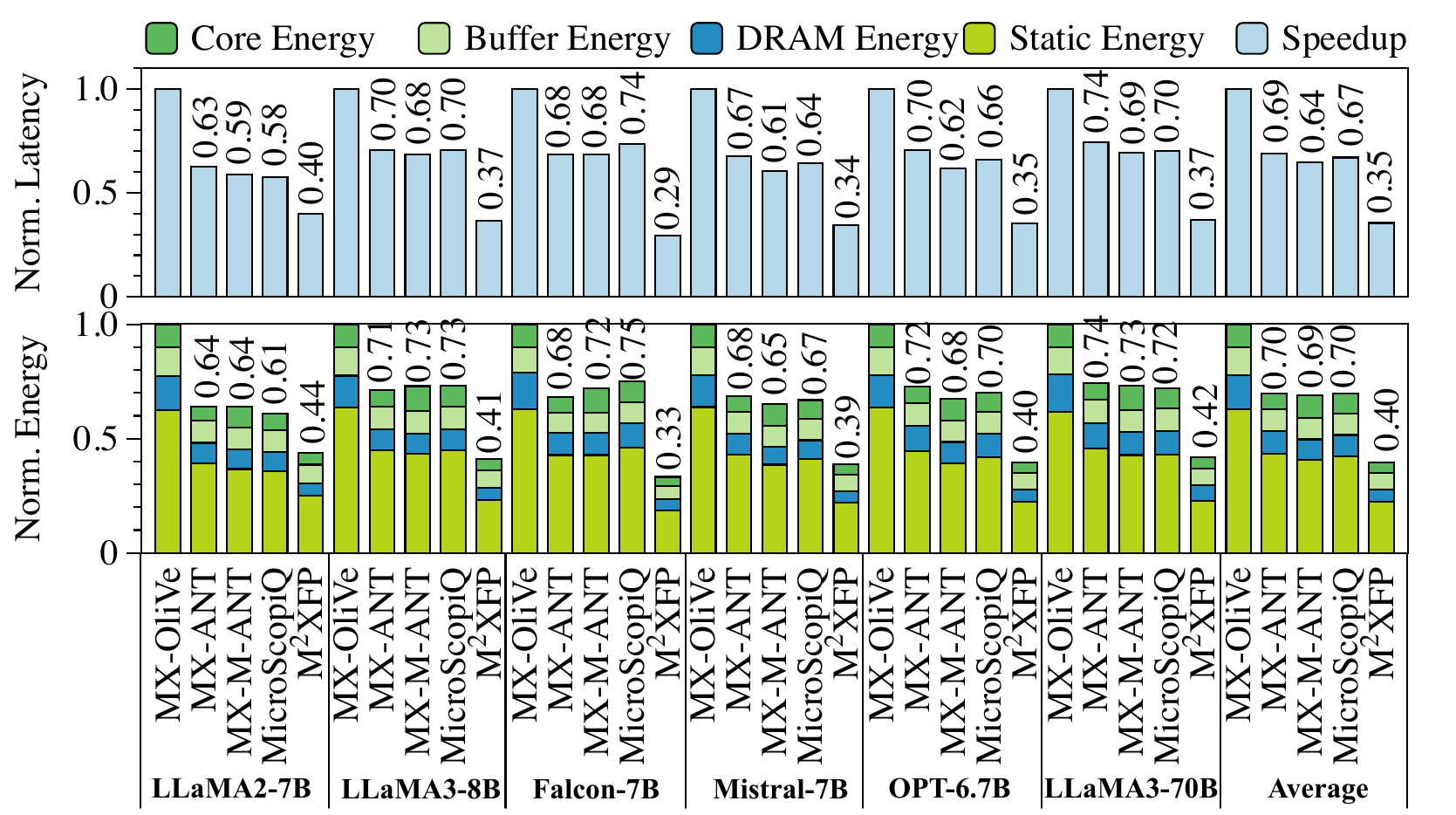}  
    \caption{The normalized latency and energy comparison between \mymethod{} and baseline accelerators.}
    \label{fig:eval_energy}
\end{figure}

\Fig{fig:eval_energy} compares performance and energy against MX-based baselines under identical systolic array sizes, differing only in decoder, encoder, or PE design.
To match accuracy, the baselines require quantizing some tensors to 8 bits, which contributes a lot to their higher latency and energy consumption.
In particular, MX-OliVe falls back to 8-bit quantization for more than 50\% of tensors, resulting in a large performance gap compared with \mymethod{}.
MX-ANT, MX-M-ANT, and MicroScopiQ achieve similar performance, but MX-M-ANT consumes extra core energy from shift-and-accumulate operations, while MicroScopiQ expends more in its ReCoN unit for outlier processing.
Overall, \mymethod{} achieves on average 1.91$\times$ speedup and 1.75$\times$ energy reduction compared to the state-of-the-art MX accelerator MicroScopiQ.

\begin{table}[t]
    \centering
    \small
    \renewcommand{\arraystretch}{1.0} 
    \caption[]{Wikitext perplexity of NVFP4 and NVFP4 with Elem-EM and Sg-EM metadata (lower is better).}
    \resizebox{0.98\columnwidth}{!}{
    \begin{tabular}{l|cccccc}
        \Xhline{1.2pt}
        \multirow{2}{*}{\textbf{Method}} & LLaMA2 & LLaMA3 & LLaMA3 & OPT & Mistral & Falcon    \\
         & 7B & 8B & 70B & 6.7B & 7B & 7B    \\
        \hline
         FP16 & 5.47 & 6.14 & 2.85 & 10.86 & 5.32 & 6.59\\
         NVFP4 & 5.81 & 7.18 & 3.63 & 11.46 & 5.76 & 6.90\\
         $\text{M}^2$-NVFP4 & 5.77 & 6.85 & 3.57 & 11.32 & 5.58 & 6.88\\
        \hline
        
        \Xhline{1.2pt}
    \end{tabular}
    }
  
    \label{eval:llm_ppl_nvfp_apply}
  \end{table}

\begin{table}[t]
    \centering
    \small
    \renewcommand{\arraystretch}{1.0} 
    \caption[]{\revise{Comparison with several algorithm schemes. The dataset is Wikitext and lll group size is 32.}}
    \resizebox{0.98\columnwidth}{!}{
    \begin{tabular}{l|ccccc}
        \Xhline{1.2pt}
        \textbf{Method} & QuaRot & DuQuant & MR-GPTQ & \mymethod{} & MR-GPTQ-\mymethod{}    \\
         \textbf{Data Type} & INT4 & INT4 & FP4 & FP4 & FP4 \\ 
        \hline
         LLaMA2-7B & 5.84 & 6.28 & 5.97 & 5.77 & 5.73  \\
         LLaMA3-8B & 7.13 & 7.90 & 7.17 & 6.84 & 6.84  \\
        \hline
        
        \Xhline{1.2pt}
    \end{tabular}
    }
  
    \label{eval:llm_algorithm}
  \end{table}

\subsection{Analysis and Discussion}

\textbf{Applying on NVFP4.}
\mymethod{} is a general design that can also extend to formats such as NVFP4.
By integrating Sg-EM for weights and Elem-EM for activations, we construct $\text{M}^2$-NVFP4 (\Tbl{eval:llm_ppl_nvfp_apply}), which yields lower perplexity than the original NVFP4.
However, since NVFP4 uses group size 16, the added metadata raises its effective bit-width from 4.5 to 5 bits.
Therefore, NVFP4 may benefit from further exploration in our encoding design framework to identify more bit-efficient designs.

\textbf{Impact of Shared Scale Calculation.}
Different ways of computing the shared scale can affect quantization error and accuracy~\cite{cook2025sixaccuratenvfp4quantization,mishra2025recipemxfp8,rtne_scale}.
We evaluate five strategies for computing the shared scale $S = 2^E$ from the block maximum $\text{amax}$ in MX-style quantization. 
The first rule, \emph{floor}, follows the OCP~\cite{ocp_mx_specification} specification and sets $E = \lfloor \log_2(\text{amax} / P) \rfloor$, where $P$ is the largest power-of-two value representable by the format (e.g., $P=4$ for FP4). 
The second rule, \emph{ceil}, instead normalizes by the maximum representable magnitude $M$ and uses $E = \lceil \log_2(\text{amax} / M) \rceil$ (e.g., $M=6$ for FP4). 
The third rule, \emph{RTN1}, replaces the ceil operation with round-to-nearest, $E = \mathrm{round}(\log_2(\text{amax} / M))$. 
The fourth rule, \emph{RTN2}, uses round-to-nearest on the normalized block maximum with respect to the largest power-of-two representable value $P$, i.e., $E = \mathrm{round}(\log_2(\text{amax} / P))$. 
Finally, the \emph{RTNE} rule, introduced in prior work~\cite{rtne_scale}, rounds $\text{amax}$ in value space, normalizes by $P$, and then applies floor to the logarithm, $E = \lfloor \log_2(\mathrm{round}(\text{amax}) / P) \rfloor$. 
Notably, for FP4 (where \(P = 4\) and \(M = 6\)), RTNE and ceil produce identical exponents because \(M = \tfrac{3}{2} P\), which ensures \(\lfloor \log_2(\mathrm{round}(a)/P) \rfloor = \lceil \log_2(a/M) \rceil\) for all block maxima \(a\); thus, the two rules are equivalent in this format.
In all cases, the shared scale is obtained as $S = 2^E$. 
Our previous experiments used only the floor rule, which corresponds to the OCP-recommended default configuration.

As shown in \Tbl{eval:llm_ceil}, for MXFP4 the ceil/RTNE rule achieves the lowest perplexity, consistent with the MXFP8 recipe~\cite{mishra2025recipemxfp8} (Appendix~A.1), whereas RTN1 performs worse because it does not address the dominant block-maximum error and introduces additional nondeterminism. 
RTN2 is close to RTNE and ceil, but is slightly worse on average.
Across all five shared-scale computation rules, \mymethod{} consistently improves accuracy over the MXFP4 baseline, indicating that its gains are robust to the choice of scaling strategy.

\begin{table}[t]
    \centering
    \small
    \renewcommand{\arraystretch}{1.0} 
    \caption[]{\revise{Wikitext perplexity of different methods to calculate the shared scale for MXFP4.}}
    \resizebox{0.7\columnwidth}{!}{
    \begin{tabular}{l|cc|cc}
        \Xhline{1.2pt}
        \multirow{2}{*}{\textbf{Models}}  & \multicolumn{2}{c|}{LLaMA2-7B} & \multicolumn{2}{c}{LLaMA3-8B}   \\
        & MXFP4 & \mymethod{}4 &  MXFP4 & \mymethod{}4  \\
        \hline
         floor & 7.15 & \textbf{5.77} & 8.30 & \textbf{6.84}   \\
         ceil/RTNE  & \textbf{6.21} & 5.80 & \textbf{7.97} & 6.96  \\
         RTN1  & 9.21 & 5.79 & 9.34 & 6.87 \\
         RTN2  & 6.26 & 5.81 & 8.08 & 7.01  \\
        \hline
        
        \Xhline{1.2pt}
    \end{tabular}
    }
  
    \label{eval:llm_ceil}
  \end{table}

\textbf{Comparison with Algorithm Schemes.}
\revise{
To demonstrate that the MXFP data format is competitive with recent algorithmic schemes~\cite{ashkboos2024quarot,lin2024duquant,egiazarian2025mrgptq}, we add comparisons with DuQuant~\cite{lin2024duquant}, QuaRot~\cite{ashkboos2024quarot} (INT), and the MX-based MR-GPTQ~\cite{egiazarian2025mrgptq}. 
Under the same group size, \mymethod{} achieves lower perplexity. Since MR-GPTQ is an algorithmic scheme and orthogonal to \mymethod{}, we also combine them; the joint gain is incremental but may improve with further tuning.
}

\textbf{Extension to Attention and KV Cache.}
While Linear layers (handling Q/K/V/O projections) dominate latency ($\sim$83\%) at typical sequence lengths of 4096, the Attention mechanism becomes significant at longer contexts, accounting for $\sim$45\% of latency at length 16384. 
Extending \mymethod{} to the KV cache is therefore crucial for sustaining its benefits across varying sequence lengths and workloads.
Furthermore, quantization strategy can be integrated with recent memory management systems~\cite{kwon2023pagedattention,xu2024vtensor,prabhu2025vattention,zheng2024sglang,guo2024gmlake} to further optimize memory usage and minimize data movement.

In practice, applying \mymethod{} to the KV cache follows the same design principles as for Linear layers. 
In Attention, K/V are both right-hand operands in GEMM ($P=QK^T$, $O=PV$). 
Systems like KIVI~\cite{liu2024kivi} and VQ-LLM~\cite{liu2025vqllm} adopt a lazy KV cache quantization policy, allowing adaptive shared scale search. 
Therefore, Sg-EM can be used for K and V, and Elem-EM for Q and P, which is compatible with \mymethod{} architecture.

\section{Conclusion}
In this paper, we presented \mymethod{}, a metadata-augmented microscaling (MX) data format that mitigates accuracy loss in 4-bit weight-activation quantization. 
We explored bit-efficient metadata allocation schemes, built a dedicated hardware unit for encoding support, and integrated it into a systolic array. 
\mymethod{} reduces accuracy loss by 70.6\% over MXFP4 and 37.3\% over NVFP4, while achieving up to 1.91$\times$ performance and 1.75$\times$ energy gains, demonstrating the practicality of metadata-driven MX formats for future LLM accelerators.

\begin{acks}
This work was supported by the National Natural Science Foundation of China (NSFC) Grants (62222210, 62532006, and 62502305), Shanghai Qi Zhi Institute Innovation Program SQZ202316, and Natural Science Foundation of Shanghai Grants (25ZR1402275).
The authors express their gratitude to the anonymous reviewers for their insightful feedback, which greatly contributed to improving this work.
We also thank our shepherd for the ongoing support and guidance during the revision process.
Any opinions, findings, and conclusions in this paper are those of the authors only and do not necessarily reflect the views of our sponsors.
\end{acks}

\bibliographystyle{ACM-Reference-Format}
\balance
\bibliography{refs}

\end{document}